\documentclass[12pt]{article}
\usepackage{amsmath}
\usepackage{graphicx}
\usepackage{enumerate}
\usepackage{natbib}
\usepackage{url} 
\usepackage{nicematrix}
\usepackage{multibib}
\usepackage{hyperref}
\usepackage{centernot}

\usepackage{tocloft}
\usepackage{etoolbox}
\usepackage{titletoc}
\usepackage{subcaption}
\usepackage{tikz}
\usepackage{enumitem}
\usepackage[lined,boxed,commentsnumbered]{algorithm2e}
\usepackage{booktabs,siunitx}

\usetikzlibrary{decorations.pathmorphing, patterns,shapes}
\DeclareRobustCommand\sampleline[1]{%
  \tikz\draw[#1] (0,0) (0,\the\dimexpr\fontdimen22\textfont2\relax)
  -- (0.99em,\the\dimexpr\fontdimen22\textfont2\relax);%
}
\newcites{supp}{Supplementary Material References}

\usepackage{amsmath,amssymb}
\usepackage{iftex}
\ifPDFTeX
  \usepackage[T1]{fontenc}
  \usepackage[utf8]{inputenc}
  \usepackage{textcomp} 
\else 
  \usepackage{unicode-math}
  \defaultfontfeatures{Scale=MatchLowercase}
  \defaultfontfeatures[\rmfamily]{Ligatures=TeX,Scale=1}
\fi
\usepackage{lmodern}
\ifPDFTeX\else  
\fi
\IfFileExists{upquote.sty}{\usepackage{upquote}}{}
\IfFileExists{microtype.sty}{
  \usepackage[]{microtype}
  \UseMicrotypeSet[protrusion]{basicmath} 
}{}
\makeatletter
\@ifundefined{KOMAClassName}{
  \IfFileExists{parskip.sty}{%
    \usepackage{parskip}
  }{
    \setlength{\parindent}{0pt}
    \setlength{\parskip}{6pt plus 2pt minus 1pt}}
}{
  \KOMAoptions{parskip=half}}
\makeatother
\usepackage{xcolor}
\makeatletter
\ifx\paragraph\undefined\else
  \let\oldparagraph\paragraph
  \renewcommand{\paragraph}{
    \@ifstar
      \xxxParagraphStar
      \xxxParagraphNoStar
  }
  \newcommand{\xxxParagraphStar}[1]{\oldparagraph*{#1}\mbox{}}
  \newcommand{\xxxParagraphNoStar}[1]{\oldparagraph{#1}\mbox{}}
\fi
\ifx\subparagraph\undefined\else
  \let\oldsubparagraph\subparagraph
  \renewcommand{\subparagraph}{
    \@ifstar
      \xxxSubParagraphStar
      \xxxSubParagraphNoStar
  }
  \newcommand{\xxxSubParagraphStar}[1]{\oldsubparagraph*{#1}\mbox{}}
  \newcommand{\xxxSubParagraphNoStar}[1]{\oldsubparagraph{#1}\mbox{}}
\fi
\makeatother

\usepackage{longtable,booktabs,array}
\usepackage{calc} 
\usepackage{etoolbox}
\makeatletter
\patchcmd\longtable{\par}{\if@noskipsec\mbox{}\fi\par}{}{}
\makeatother
\IfFileExists{footnotehyper.sty}{\usepackage{footnotehyper}}{\usepackage{footnote}}
\makesavenoteenv{longtable}
\usepackage{graphicx}
\makeatletter
\def\maxwidth{\ifdim\Gin@nat@width>\linewidth\linewidth\else\Gin@nat@width\fi}
\def\maxheight{\ifdim\Gin@nat@height>\textheight\textheight\else\Gin@nat@height\fi}
\makeatother
\setkeys{Gin}{width=\maxwidth,height=\maxheight,keepaspectratio}
\makeatletter
\def\fps@figure{htbp}
\makeatother

\makeatletter
\@ifpackageloaded{caption}{}{\usepackage{caption}}
\AtBeginDocument{%
\ifdefined\contentsname
  \renewcommand*\contentsname{Table of contents}
\else
  \newcommand\contentsname{Table of contents}
\fi
\ifdefined\listfigurename
  \renewcommand*\listfigurename{List of Figures}
\else
  \newcommand\listfigurename{List of Figures}
\fi
\ifdefined\listtablename
  \renewcommand*\listtablename{List of Tables}
\else
  \newcommand\listtablename{List of Tables}
\fi
\ifdefined\figurename
  \renewcommand*\figurename{Figure}
\else
  \newcommand\figurename{Figure}
\fi
\ifdefined\tablename
  \renewcommand*\tablename{Table}
\else
  \newcommand\tablename{Table}
\fi
}
\@ifpackageloaded{float}{}{\usepackage{float}}
\floatstyle{ruled}
\@ifundefined{c@chapter}{\newfloat{codelisting}{h}{lop}}{\newfloat{codelisting}{h}{lop}[chapter]}
\floatname{codelisting}{Listing}

\makeatother
\makeatletter
\@ifpackageloaded{caption}{}{\usepackage{caption}}
\@ifpackageloaded{subcaption}{}{\usepackage{subcaption}}
\makeatother

\ifLuaTeX
  \usepackage{selnolig}  
\fi
\usepackage[]{natbib}
\usepackage{bookmark}

\IfFileExists{xurl.sty}{\usepackage{xurl}}{} 
\usepackage{xr}

\newcommand{\anon}{1}

\usepackage[mathscr]{eucal}

\usepackage{dlfltxbcodetips,bbm,mathtools,amsfonts,amsmath,amssymb,fancybox,graphicx,bm,latexsym}

\newcommand{\actorset}{\mathscr{N}}
\newcommand{\timeset}{\mathscr{T}}
\newcommand{\dyadset}{\mathscr{D}}
\newcommand{\actorsubset}{\mathscr{D}}
\newcommand{\baselinechangeset}{\mathscr{B}}
\newcommand{\eventsoccurset}{\mathscr{U}}
\usepackage{tikz}
\usetikzlibrary{bayesnet}
\usetikzlibrary{fit,positioning}

\newcommand{\spectralnormvec}[1]{{|\!| #1 |\!|_2}}

\newcommand{\zeronormvec}[1]{{\left|\!\left| #1 \right|\!\right|_1}}
\newcounter{proof}

\usepackage[tikz]{bclogo}

\newcommand{\bt}{
 \begin{bclogo}[logo= \bcloupe 
,couleur={rgb:orange,0;yellow,0;white,1},arrondi=0.1,ombre=false]}
	\newcommand{\et}{\end{bclogo}\s}
\newcommand{\btt}{\begin{box}}
	\newcommand{\ett}{\end{box}}

\newcommand{\btheorem}{\begin{bclogo}[couleur={rgb:orange,0;yellow,0;white,1},arrondi=0.1,logo=\bcplume,ombre=true]{Theorem}}
	\newcommand{\ettheorem}{\end{bclogo}}

\newcommand{\bsh}{\begin{bclogo}[couleur={rgb:orange,0;yellow,0;white,1},arrondi=0.1,logo=\bcpanchant,ombre=true]}
	\newcommand{\esh}{\end{bclogo}}

\newcommand{\benum}{\begin{enumerate}}
	\newcommand{\eenum}{\end{enumerate}}

\newcommand{\bq}{\begin{quote}\em}
	\newcommand{\eq}{\end{quote}}
\newcommand{\bbq}{\begin{quote}\bf\em}
	\newcommand{\ebq}{\end{quote}}

\newcommand{\hide}[1]{}
\newcommand{\ghost}[1]{}
\newcommand{\ba}{\begin{array}{llllllllll}}
	\newcommand{\ea}{\end{array}}
\newcommand{\bea}{\begin{equation}\begin{array}{llllllllll}}
		\newcommand{\eea}{\end{array}\end{equation}}

\newcommand{\beno}{\begin{equation}\begin{array}{llllllllll}\nonumber}
		\newcommand{\be}{\begin{equation}\begin{array}{llllllllll}}
				\newcommand{\ee}{\end{array}\end{equation}}
		\newcommand{\bcc}{\begin{equation}\begin{array}{cccccccccc}}
				\newcommand{\ecc}{\end{array}\end{equation}}
		\newcommand{\bi}{\begin{itemize}}
			\newcommand{\ei}{\end{itemize}}
		\newcommand{\ben}{\begin{enumerate}}
			\newcommand{\een}{\end{enumerate}}
\newcommand{\alert}[1]{\textcolor{red}{\bf{#1}}}

		\newcommand{\dsum}{\displaystyle\sum\limits}
		\newcommand{\dint}{\displaystyle\int\limits}

		\newcommand{\s}{\vspace{0.25cm}}

		\newcounter{comment}

		\newcounter{example}

		\newcounter{counterexample}

		\newcounter{definition}
		
		\newcounter{theorem}

		\newcounter{proposition}
		\newenvironment{proposition}[1][]{\refstepcounter{proposition}\par\medskip\indent%
			\textbf{Proposition~\theproposition #1}. \rmfamily}{\medskip}
		
		\newcounter{result}

		\newcounter{tproof}
		\newcounter{corollary}

		\newcounter{cproof}
		\newcounter{lemma}

		\newcounter{com}

		\newcounter{lproof}
		\newcounter{assumption}

		\newif\ifmydraft
		\mydraftfalse

		\ifmydraft
		
		\pagecolor{black!20}
		\else
		
		\fi

		\ifmydraft
		
		\pagecolor{black!20}
		\fi
		
		\usepackage{multicol}

		\makeatletter
		\newcommand*{\deq}{\mathrel{\rlap{%
					\raisebox{0.3ex}{$\m@th\cdot$}}%
				\raisebox{-0.3ex}{$\m@th\cdot$}}=}
		\makeatother

\begin{document}

\def\spacingset#1{\renewcommand{\baselinestretch}%
{#1}\small\normalsize} \spacingset{1}


\if1\anon
{
  \title{\bf Scalable Durational Event Models: Application to Physical and Digital Interactions}
  \author{
  Cornelius Fritz$^1\footnote{\textbf{Corresponding author:} \href{mailto:fritzc@tcd.ie}{fritzc@tcd.ie}.
}$, Riccardo Rastelli$^{2,3}\footnote{
Riccardo Rastelli, Michael Fop, and Alberto Caimo contributed equally to the work. Alberto Caimo conceived the original idea of the model.
}$, Michael Fop$^2\footnotemark[2]$,  Alberto Caimo$^2\footnotemark[2]$\\[.2cm] $^1$School of Computer Science and Statistics, Trinity College Dublin  \vspace{.1cm}\\ $^2$School of Mathematics and Statistics, University College Dublin \\
$^3$Rinn Artificial Intelligence, University College Dublin, Ireland
  }
  \maketitle
} \fi

\if0\anon
{
  \bigskip
  \bigskip
  \bigskip
  \begin{center}
    {\LARGE\bf Scalable Durational Event Models: Application to Physical and Digital Interactions}
\end{center}
  \medskip
} \fi

\bigskip

\begin{abstract}
Durable interactions are increasingly observed in social network analysis with precise timestamps.
Phone and video calls, for instance, are events to which a specific duration can be assigned.
We term data encoding interactions with start and end times ``durational event data''.
Recent advances in data collection have enabled the observation of such data over extended periods and across large populations of actors. 
Methodologically, we propose the Durational Event Model, an extension of Relational Event Models that decouples the modeling of event incidence from event duration.
Computationally, we derive a fast, memory-efficient, and exact block-coordinate ascent algorithm to facilitate large-scale inference.
Theoretical complexity analysis and numerical simulations demonstrate the computational superiority of this approach over state-of-the-art methods.
We apply the model implemented in the R package \texttt{redeem} to physical and digital interactions among college students in Copenhagen. 
Our empirical findings reveal that past interactions drive physical interactions, whereas digital interactions are influenced predominantly by friendship ties and prior dyadic contact. 
\end{abstract}
\noindent%
{\it Keywords:} Block coordinate algorithms,  Dynamic networks, Large event data, Minorization-maximization algorithms, Relational event model
\vfill

\newpage
\spacingset{1.8} 

\setlist[itemize]{label=\raisebox{0.3ex}{\scalebox{0.6}{$\bullet$}}}

\renewcommand{\thefootnote}{\arabic{footnote}}

\section{Introduction}
\label{sec:intro}

\hide{
Dynamic network models have historically assumed that edges are temporally extensive, in that they are present over some time \citep{butts_relational_2017}, and that continuous observations of the changes in the network are available \citep{holland_dynamic_1977}. 
However, 
early empirical applications were constrained since networks were measured via surveys, making it challenging to fully make use of dynamic models. 
Fast forward thirty years, and 
\citet[p. 1]{savage_coming_2007} argue that ``both the sample survey and the in-depth interview are increasingly dated research methods, which are unlikely to provide a robust base for the jurisdiction of empirical sociologist in the coming decades''. 
In response, \cite{beer_sociology_2007} suggest using \textsl{transactional} data that comprises of routinely gathered information and social media data. 
For network data, this led to large-scale online networks growing rapidly due to rapid digitization \citep{lazer_computational_2009}. 
Contrasting the few hundred actors in a network measured through surveys or interviews, digitally measured networks encompass millions of actors and edges and are cheap to collect \citep{wagnerMeasuringAlgorithmicallyInfused2021a}. 
}

Driven by rapid digitization, the availability of large-scale online networks is growing at a fast pace \citep{lazer_computational_2009}. 
In contrast to traditional sociometric studies, which typically span a few hundred actors in a network, digitally measured networks encompass thousands of actors and edges and are relatively easy to collect \citep{wagnerMeasuringAlgorithmicallyInfused2021a}. 
The raw observations underlying these large networks are often automatic logs with precise timestamps of events; for instance, the time an email was sent. 
Complementing these digital networks, wearable sociometric badges allow continuous measurements of physical interactions \citep{eagle_reality_2006}. 
These data encode massive networks representing relations in contexts ranging from hospital patient transfers to instant messaging platforms.

Such large-scale networks pose distinct computational and inferential challenges.
Most algorithms for estimating models for network data are based on Markov chain Monte Carlo (MCMC) algorithms \citep{hummelImprovingSimulationbasedAlgorithms2012}, scaling poorly to high-dimensional parameters and larger networks. 
Since a growing network warrants an increasingly complex model, this error-prone regime is expected for large networks.
One way to mitigate these issues of larger networks is to base inference on an approximate likelihood \citep{rafteryFastInferenceLatent2012} or online algorithms \citep{corneck2025}.
In settings with latent variables, variational approximations are often employed, turning sampling from a posterior via MCMC into an optimization problem. 
For the resulting optimization problem, Minorization-Maximization (MM) algorithms were devised to obtain robust and scalable solutions \citep{vu2013}. 

Often, one can represent these networks as sequences of events that are observed continuously in the time interval $\timeset \coloneqq [0,T]$ between actors in the set $\actorset \coloneqq \{1,\ldots, N\}$.
Relational event models (REMs, \citealp{butts_relational_2008}) provide a framework for analyzing such events by specifying a joint stochastic process $\bm{N}(t) = (N_{i,j}(t))$ that counts all possible events up to time $t$.
For each pair of actors $(i,j) \,\in\, \dyadset \coloneqq \{(i,j) \in \actorset \times \actorset:  i<j\}$ at time $t\in \timeset$, the counting process $N_{i,j}(t)$ is governed by a dyadic intensity, which is a function of the history up to but not including time $t$, denoted by the $\sigma$-algebra $\mathscr{H}_t = \sigma(\{\bm{N}(u): u < t\})$: 
\be
\label{eq:intensity_rem}
      \lambda_{i,j}\left(t\mid \mathscr{H}_t, \bm{\theta}\right) &=& g\left(\bm{\theta}^\top \bm{s}_{i,j}(\mathscr{H}_t)\right). 
\ee
The summary statistics $\bm{s}_{i,j}(\mathscr{H}_t) \in \mathbb{R}^p$ for actors $i$ and $j$ are functions of $\mathscr{H}_t$.
An entry of $\bm{s}_{i,j}(\mathscr{H}_t)$ might be the count of common partners between actors $i$ and $j$ until $t$. 
These statistics are weighted by the parameter $\bm{\theta}\in \mathbb{R}^p$. 
The function $g$ is a non-decreasing twice differentiable function mapping to $\mathbb{R}^+$, which is commonly the  identity, $g: x\mapsto x$ (for positive $x$), or exponential function $g: x\mapsto \exp(x)$, depending on the sufficient statistics affecting the dyadic intensity additively or multiplicatively. 
With $\bm{s}_{i,j}(\mathscr{H}_t) = (1,\int_0^t \exp(-(t-u))\, \text{d}\, N_{i,j}(u))^\top$, the former case encompasses Hawkes processes
\citep{passino2023}. 
In the latter case, the model corresponds to the original formulation of \citet{butts_relational_2008}. 
This model class is widely used in social science (see, e.g.,\citealp{kitts2017}) and was extended to account for spurious events \citep{fritz_all_2023} and uncover latent communities \citep{matias2018} to name a few. 
To manage the large size of event data, caching \citep{vu_continuous-time_2011} and sampling-based procedures  \citep{lerner2020} were introduced.

Although timestamps are often available for durable ties, relational event modeling has so far almost exclusively focused on events without any duration. 
Many time-dependent interactions, such as phone or Zoom calls, naturally have a temporal duration associated with each observed event. 
We introduce the term ``durational event'' to describe this particular type of event. 
Differentiating between duration and incidence is crucial for understanding the strength and dynamics of interactions.
While duration captures interaction depth and time investment, engagement and volume are reflected in frequency of occurrences, offering complementary insights. 
One way to model such data is to treat the duration of an event as an attached weight or mark and use a model for weighted events  \citep{lerner_modeling_2013}. 
However, this representation disregards that the duration of an interaction is an endogenous process, influenced by factors that arise during the interaction itself or by external factors, rather than being determined at the start of the interaction. 
\hide{
Modeling these aspects separately provides a more realistic framework than a marked counting process, where duration could be treated as a random mark \citep{lerner_modeling_2013}. 
}

\citet{stadtfeld_dynamic_2017} took first steps toward analyzing a related data structure by proposing a model for coordination ties. 
These ties are a particular case of durational events, where the creation of a link between two actors is a two-sided decision process. 
Still, their application focused solely on the incidence of events, disregarding their duration.
\citet{hoffman_model_2020} introduced a model for group-based interactions, where actors can join and leave groups and \citet{rastelli_stochastic_2020} propose a stochastic block model for durational events.  

Despite these initial steps, there is a gap in the literature on how to model general durational events beyond the discussed special cases.
We address this limitation by introducing the Durational Event Model (DEM) as a general framework for analyzing durational events in Section \ref{sec:dem}. 
The DEM is related to the Separable Temporal Exponential Random Graph Model (STERGM, \citealp{krivitsky_separable_2014}) and Stochastic Actor-Oriented Models with gratification functions \citep{SnDu97}, which similarly model tie formation and dissolution. 
However, our framework assumes continuous observations, while the mentioned approaches model snapshots of networks at discrete times. 
The DEM faces the aforementioned challenges of large-scale networks, since its number of parameters grows with the number of actors and length of the observed time-frame. 
Thus, standard techniques, such as the Newton-Raphson method,  are impractical for estimation in most applications. 
We develop in Section \ref{sec:computing} a block-coordinate ascent method based on separate minorization-maximization and closed-form steps to overcome this limitation. 
In Section \ref{sec:simulation_study}, we assess the performance of our algorithm in a simulation study. 
Next, we apply our model class to physical and digital interaction data from the Copenhagen Networks Study \citep{sapiezynskiInteractionDataCopenhagen2019} 
in Section \ref{sec:applications}.  
Finally, we discuss possible future extensions in Section \ref{sec:discussion}. 
Our method is implemented in the $\mathtt{redeem}$ package for \texttt{R} available on CRAN \citep{fritz2026b}.

\section{Durational Event Model}
\label{sec:dem}

A durational event between the pair of actors $(i,j) \,\in\, \dyadset$, beginning at time $t_b \in \timeset$ and ending at time $t_e \in \timeset$, with $t_b<t_e$, is represented by the four-dimensional tuple $d = (i,\,j,\,t_b,\,t_e)$.
Henceforth, we focus on undirected durational events, although the extension to directed events is straightforward.
We define by $\actorsubset^{0 \rightarrow 1}(t) \subseteq \dyadset$ the set of possible actor pairs that may experience the start of a durational event at $t \in \timeset$ (i.e., change from a ``not interacting'' status, denoted with 0, to an ``interacting'' status, denoted with 1). 
Similarly, $\actorsubset^{1 \rightarrow 0}(t) \subseteq \dyadset$ is the set of possible actor pairs that may experience the end of a durational event at $t \in \timeset$.
We constrain these sets so that $\actorsubset^{0 \rightarrow 1}(t) \, \cap \, \actorsubset^{1 \rightarrow 0}(t) = \emptyset$ holds, to guarantee that an interaction between a pair cannot start and end at the same time. 
Then $\actorsubset^{0 \rightarrow 1}(t) \, \cup \, \actorsubset^{1 \rightarrow 0}(t) \coloneqq \actorsubset^\star(t)$ denotes the set of dyads where an event can be observed at time $t \in \timeset$.
To account for sparse interaction, one can define these sets to pick which dyads are governed by the specified model \citep{kreis_nonparametric_2019, matias2018}. 
If durational events correspond to phone calls, actors may only engage in one interaction at a time, which is represented by a constraint on these sets: 
observing call event $(i,j,t_b,t_e)$, implies $(i,h)$ and $(j,h)\notin \actorsubset^\star(u)$ for $u \in [t_b,t_e)$ and $h \neq i,j$. 
We assume that these sets are known for $t \in \timeset$.
\hide{
We accommodate this setting by letting the set of possible pairs of actors that may experience the start and end of a durational event at $t \in \timeset$ by $\actorsubset^{0 \rightarrow 1}(t) \subseteq \dyadset$ and $\actorsubset^{1 \rightarrow 0}(t) \subseteq \dyadset$,  respectively.
}

\subsection{Model Specification}
\label{sec:model_specification}

To model the dynamics of durational events, we specify the DEM via two separate counting processes: the formation process, $N_{i,j}^{0 \rightarrow 1}(t)$, counting the number of times that $i$ and $j$ have started an interaction up to time point $t$; and the dissolution process, $N_{i,j}^{1 \rightarrow 0}(t)$, 
counting the number of times that the actors have stopped interacting before and up to $t$. 
Together, the two stochastic processes $\bm{N}^{0\rightarrow 1}(t) = (N_{i,j}^{0 \rightarrow 1}(t))$ and $\bm{N}^{1\rightarrow 0}(t) = (N_{i,j}^{1 \rightarrow 0}(t))$ count the frequency with which all actor pairs transition between the states ``not interacting'' and ``interacting'' until time $t\in \timeset$.
We characterize these processes by their respective instantaneous probabilities of a jump, called the incidence intensity, $\lambda_{i,j}^{0 \rightarrow 1}(t\mid \mathscr{H}_t)$, and the duration intensity, $\lambda_{i,j}^{1 \rightarrow 0}(t\mid \mathscr{H}_t)$. 


The incidence and duration intensities between actors $i$ and $j$ in $\actorset$ in the DEM may be influenced by their previous interactions, past interactions involving common neighbors, and ongoing interactions within the broader population. 
We adopt the multiplicative form with $g: x\mapsto \exp(x)$ from \eqref{eq:intensity_rem} with a self-excitation mechanism based on occurrences. 
This contrasts with Hawkes processes, which usually have an additive form with self-excitation driven by the length of time since previous events. 
As such, our work is more aligned with established models in the social sciences \citep{butts_relational_2008}, and computationally tractable due to piecewise-constant intensities, upon which our algorithm from Section \ref{sec:computing} relies. 
The intensities $\lambda_{i,j}^{0 \rightarrow 1}(t)$ and $\lambda_{i,j}^{1 \rightarrow 0}(t)$ are $\mathscr{H}_t$-predictable functions of the observed process history:
\be
\label{eq:intensity_0_1}
       \lambda_{i,j}^{0 \rightarrow 1}\left(t\mid \mathscr{H}_t, \bm{\theta}^{0 \rightarrow 1}\right) &=& \exp\left(\bm{\alpha}^{0 \rightarrow 1}\, \bm{s}_{i,j}^{0 \rightarrow 1}(\mathscr{H}_t) + \beta^{0 \rightarrow 1}_i +\beta^{0 \rightarrow 1}_j + f(t, \bm \gamma^{\,0 \rightarrow 1})\right)\\
      \lambda_{i,j}^{1 \rightarrow 0}\left(t\mid \mathscr{H}_t, \bm{\theta}^{1 \rightarrow 0}\right) &=& \exp\left(\bm{\alpha}^{1 \rightarrow 0} \,\bm{s}_{i,j}^{1 \rightarrow 0}(\mathscr{H}_t) + \beta^{1 \rightarrow 0}_i +\beta^{1 \rightarrow 0}_j + f(t,\bm{\gamma}^{\,1 \rightarrow 0})\right),
\ee
for $(i,j) \in \actorsubset^{0 \rightarrow 1}(t)$ or $\actorsubset^{1 \rightarrow 0}(t)$, respectively, where 
\begin{itemize} 
    \item 
    $\bm{s}_{i,j}^{0 \rightarrow 1}(\mathscr{H}_t) = (s_{i,j,1}^{0 \rightarrow 1}(\mathscr{H}_t), \ldots, s_{i,j,P}^{0 \rightarrow 1}(\mathscr{H}_t))^\top \in \mathbb{R}^P$ are summary statistics for the pair of actors $(i,j)$ that are functions of the history $\mathscr{H}_t= \sigma(\{\bm{N}^{0 \rightarrow 1}(u), \bm{N}^{1 \rightarrow 0}(u): u < t\})$. 
    These summary statistics capture dependencies between coordinates of $\bm{N}^{0 \rightarrow 1}(t)$ and $\bm{N}^{1 \rightarrow 0}(t)$ by taking into account endogenous and exogenous processes. 
    Endogenous processes originate from events being modeled, such as the number of common partners actors $i$ and $j$ had up to $t$, while exogenous processes encompass any covariate-driven processes that, for example, capture homophily based on gender.
    More details on and examples of these statistics are provided in Section \ref{sec:summary_statistics}.
    \item $\bm{\alpha}^{0 \rightarrow 1} = (\alpha^{0 \rightarrow 1}_1, \ldots, \alpha^{0 \rightarrow 1}_P)\in \mathbb{R}^{1\times P}$ are parameters corresponding to $\bm{s}_{i,j}^{0 \rightarrow 1}(\mathscr{H}_t)$.
    \item $\bm{\beta}^{0 \rightarrow 1} = ({\beta}^{0 \rightarrow 1}_1, \ldots, {\beta}^{0 \rightarrow 1}_N)^\top \in \mathbb{R}^N$ are popularity parameters accounting for different overall activity levels of each actor. 
    \item 
$f: \timeset \times \mathbb{R}^Q \mapsto \mathbb{R}$ with $f(t,\bm{\gamma}^{0\rightarrow 1}) = \sum_{q = 1}^{Q} \gamma_q^{0\rightarrow 1} \, \mathbb{I}(c_{q-1}\leq t < c_q)$ is a step-function with user-defined intervals capturing temporal variations in the data. 
    The indicator function  $\mathbb{I}(c_{q-1} \leq t < c_q)$  takes the value 1 if  $c_{q-1} \leq t < c_q$  and 0 otherwise, while the parameter vector $\bm{\gamma}^{0 \rightarrow 1} = (\gamma^{0 \rightarrow 1}_1, \ldots, \gamma^{0 \rightarrow 1}_Q)\in \mathbb{R}^Q$  determines the value of  $f$  within the $Q$ segments.
      To ensure identifiability, we impose $\gamma_1^{0\rightarrow 1} = 0$. 
      Section \ref{sec:applications} and Supplement B.5  provide concrete examples and sensitivity analyses for specifying this step-function.
\end{itemize}
We collect all parameters in
$\bm{\theta}^{0 \rightarrow 1} \coloneqq ((\bm{\alpha}^{0 \rightarrow 1})^\top, \bm{\beta}^{0 \rightarrow 1}, \bm{\gamma}^{\,0 \rightarrow 1})^\top$, 
with analog quantities defined for the {duration} intensity. 
Both intensities are piecewise constant functions, which may change at two types of time points:
the baseline change points, $\baselinechangeset\coloneqq \{c_1, \ldots, c_Q\}$,  and the time points where events occur or the observational window ends, $\eventsoccurset \coloneqq\{t_1, \ldots, t_M, T\}$. 
Here, $Q$ is the number of time points where the baseline intensity changes and $M$ denotes the number of observed durational events.
Therefore, conditional on $\mathscr{H}_{\tilde t}$, $N_{i,j}^{0 \rightarrow 1}(t)$ and $N_{i,j}^{1 \rightarrow 0}(t)$ for $(i,j) \in \dyadset$ behave like homogeneous Poisson processes within each interval with constant intensity, $(\tilde t, t]$.
Setting $\lambda_{i,j}^{1 \rightarrow 0}\left(t|\mathscr{H}_t, \bm{\theta}^{1 \rightarrow 0}\right) = \infty$ for $(i,j) \in \actorsubset^{1 \rightarrow 0}(t)$ and $t\in\timeset$, shows that the REM of \eqref{eq:intensity_rem} is a special case of \eqref{eq:intensity_0_1}. 


\hide{
\begin{figure}[t!]
    \centering
    \includegraphics[width=0.5\textwidth]{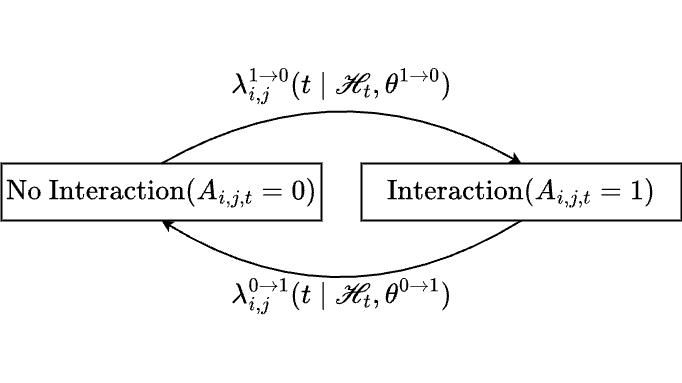}
    \caption{Graphical illustration of the two states of the network edges for durational events.}
    \label{fig:two_stage}
\end{figure}
}
The observed durational events define an instantaneous adjacency matrix $\bm{a}(t) \in \{0,1\}^{N\times N} = (a_{i,j}(t))$, where $a_{i,j}(t) \in \{0,1\}$ indicates the presence ($a_{i,j}(t) = 1$) or absence ($a_{i,j}(t) = 0$) of an interaction of actor pair $(i,j) \in \dyadset$ at time $t \in \timeset$. 
Observing durational event $d = (i,\,j,\,t_b,\,t_e)$ implies $a_{i,j}(t) = 1$ for $t \in [t_b,t_e]$, so that the instantaneous intensity of a state change for dyad $(i,j)$ is given by the mixture of both intensities defined in \eqref{eq:intensity_0_1}:
\be
\label{eq:intensity_comb}
\lambda_{i,j}\left(t\mid \mathscr{H}_t, \bm{\theta}\right) \,=\, \lambda_{i,j}^{0 \rightarrow 1}\left(t\mid \mathscr{H}_t, \bm{\theta}^{0 \rightarrow 1}\right)\, (1-a_{i,j}(t)) +\lambda_{i,j}^{1 \rightarrow 0}\left(t\mid \mathscr{H}_t, \bm{\theta}^{1 \rightarrow 0}\right)\, a_{i,j}(t). 
\ee 


\subsection{Summary Statistics}
\label{sec:summary_statistics}
The statistics $\bm{s}_{i,j}^{\,0 \rightarrow 1}(\mathscr{H}_t)$ and $\bm{s}_{i,j}^{\,1 \rightarrow 0}(\mathscr{H}_t)$ formalize the influence of past interactions on the likelihood of future events. 
Careful consideration is necessary regarding how this influence is specified. 
\citet{aalen2007} note that a valid counting process must have an intensity function that remains finite at all times to prevent an explosion where the intensity diverges. 
The Feller criterion \citep[Sect. 3.2.2]{aalen2007} guarantees non-explosion if the intensity function satisfies a linear growth bound.
Applying the transformation $x \mapsto \log(x+1)$ to past event counts in the summary statistics, ensures that the intensities in \eqref{eq:intensity_0_1} satisfy this condition.

\begin{figure}[t!]
    \centering
    \includegraphics[width=0.8\textwidth, page = 4]{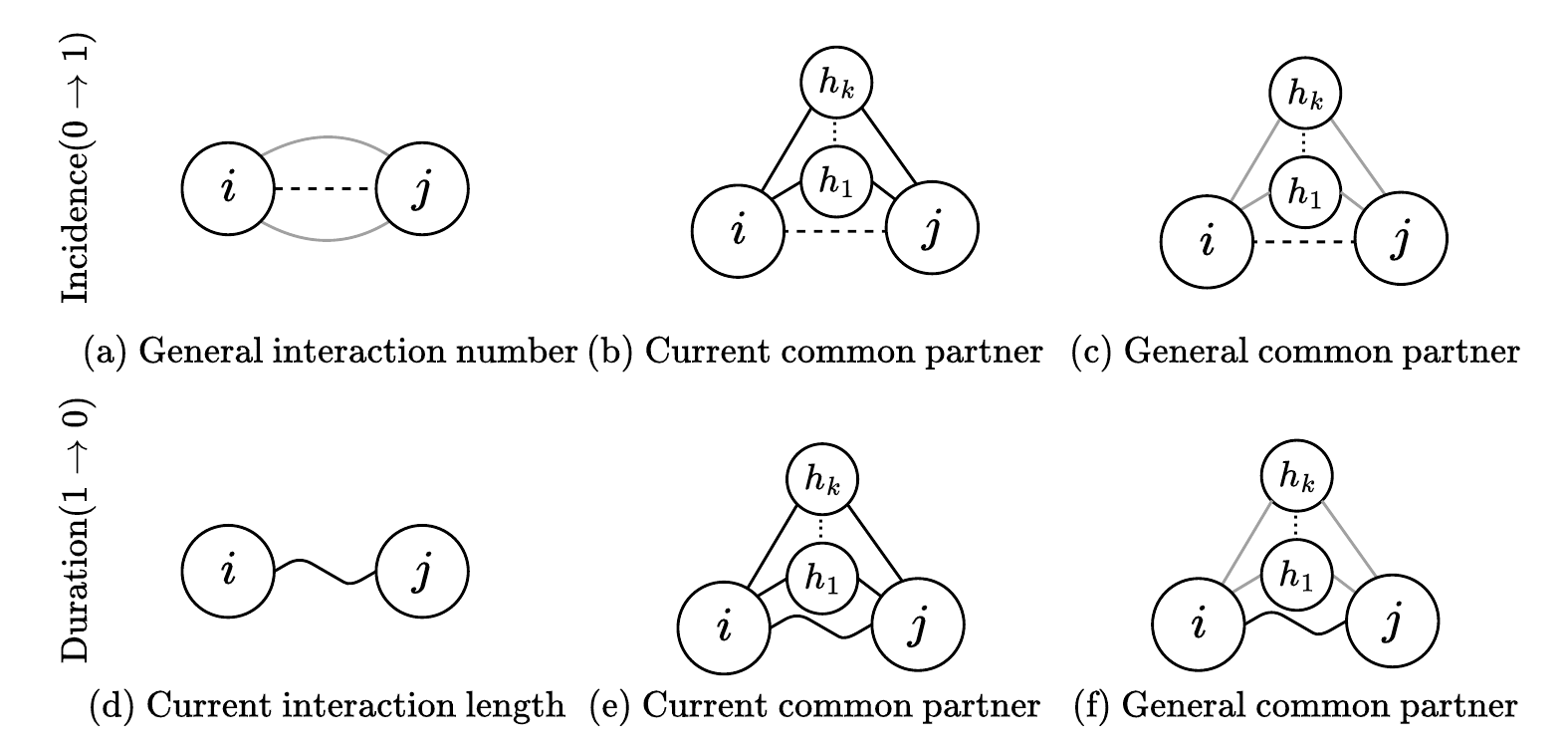}
     \caption{Graphs illustrating the proposed summary statistics. Dashed lines (\sampleline{dashed, thick}) refer to transitions from $0 \rightarrow 1$, while the wiggly line (\sampleline{decoration=snake,decorate, thick}) relates to $1 \rightarrow 0$. Other observed interactions are drawn as solid lines, in black if they are currently interacting actors (\sampleline{thick}) and gray if the event occurred sometime in the past (\sampleline{lightgray, thick}).}
    \label{fig:suff_stats}
\end{figure}

With two further exceptions, the summary statistics can then be defined similarly to their counterparts in REMs: 
First, we assume that the incidence and duration models rely on distinct sets of statistics. 
Below, we introduce statistics which can be incorporated into either $\bm{s}_{i,j}^{0 \rightarrow 1}(\mathscr{H}_t), \bm{s}_{i,j}^{1 \rightarrow 0}(\mathscr{H}_t) $, or both. 
Second, the statistics can leverage additional information from durational events, such as which actors are currently interacting. 
With $t^-$ being the time right before $t$, let $v_{i,j}(t^-)$ indicate whether they have interacted before time $t^-$.
The elapsed time since actors $i$ and $j$ started their interaction to time $t^-$ is $\Delta_{i,j}(t^-)$.
Given these quantities, we provide a list of exemplary summary statistics, illustrated      in Figure \ref{fig:suff_stats}: 
\begin{itemize}
    \item $s_{i,j,CCP}(\mathscr{H}_t) \,=\, \log\left(\sum_{h \notin \{i,j\}} a_{i,h}(t^-)\, a_{h,j}(t^-)+1\right)$: statistic representing the number of currently active partners shared by $i$ and $j$ at time $t;$
    \item 
    $s_{i,j,GCP}(\mathscr{H}_t) \,=\, \log\left(\sum_{h \notin \{i,j\}} v_{i,h}(t^-)\, v_{h,j}(t^-)+1\right)$: statistic representing the number of general common partners of $i$ and $j$ up to time $t;$
    \item $s_{i,j,NI}(\mathscr{H}_t) \,=\,  \log(N_{i,j}(t^-)+1)$: statistic representing the cumulative sum of interactions between $i$ and $j$ up to $t^-$ ($N_{i,j}(t^-)$ can be $N_{i,j}^{0\rightarrow 1}(t^-)$ or $N^{1\rightarrow 0}_{i,j}(t^-)$);
   \item $s_{i,j,IL}^{\,1\rightarrow 0}(\mathscr{H}_t) \,=\,  \log(\Delta_{i,j}(t^\star)+1)$: 
     interaction length statistic representing the duration of the current interaction between actors $i$ and $j$ until time $t^\star$,  denoting the most recent time point before $t$ an event was observed;
    \item $s_{i,j,\bm z}(\mathscr{H}_t) \,=\,  z_{i,j}$ : dyadic covariate effect, where, e.g.,  $z_{i,j} = |x_{i,1} - x_{j,1}|$ or $ z_{i,j} = \mathbb{I}(x_{i,2} = x_{j,2})$ depending on whether the effect is based on a categorical, $\bm x_1 = (x_{1,1}, \ldots, x_{N,1})$, or continuous, $\bm x_2 = (x_{1,2}, \ldots, x_{N,2})$, exogenous covariate.
\end{itemize}
\hide{
These statistics illustrate two types of dependence in our framework: intra-dyadic dependence, the dependence between counting processes $N_{i,j}^{0\rightarrow 1}(t)$ and $N_{i,j}^{1\rightarrow 0}(t)$, and 
cross-dyadic dependence, the dependence between counting processes $N_{i,j}^{0\rightarrow 1}(t)$ and $N_{i,h}^{0\rightarrow 1}(t)$, via the shared history $\mathscr{H}_t$. 
For the former, incorporating $s_{i,j,GCP}^{1\rightarrow0}(\mathscr{H}_t)$ shows that 
}
Parallels between the DEM and proportional hazard model facilitate interpreting the estimated coefficients.
Consider the incidence intensities of events $d_1 = (i,j,t,t_e)$ and $d_2 = (h,k,t,t_e)$, where all summary statistics are identical with the exception that the $p$th entry of the summary statistics for $d_1$ is one unit higher than for $d_2$ (i.e., $s_{i,j,p}(\mathscr{H}_t) = s_{h,k,p}(\mathscr{H}_t) + 1$). 
The relationship between their intensities to start an interaction at time $t$ is: 
\be
\label{eq:interpretation_exp}
\lambda_{i,j}^{0 \rightarrow 1}\left(t \mid \mathscr{H}_t, \bm{\theta}^{0 \rightarrow 1}\right) &=& \exp(\alpha_p^{0 \rightarrow 1})\lambda_{h,k}^{0 \rightarrow 1}\left(t\mid \mathscr{H}_t, \bm{\theta}^{0 \rightarrow 1}\right).
\ee
Put differently, with $\alpha_p^{0 \rightarrow 1} > 0$, observing $d_1$ is $\exp(\alpha_p^{0 \rightarrow 1})$ times more likely than $d_2$.

We log-transform statistics involving counts of past events, such as the number of current common partners between actors $i$ and $j$ until time $t$. 
Similar to geometrically weighted statistics in the context of ERGMs \citep{hunter2007}, this transformation formalizes the intuition that the initial change in a statistic has the greatest impact with subsequent changes having diminishing returns.
A change from $l$ to $l+1$ on the original scale of a $\log(\,\cdot\; +1)$-transformed statistic affects the intensity by the multiplicative factor $\left({(l+2)}/{(l+1)}\right)^{\alpha_p^{0 \rightarrow 1}}$. 
Setting $l = 0$, the quantity $2^{\alpha_p^{0 \rightarrow 1}}$ represents the multiplicative effect of the first unit increase of a statistic on its original scale and is therefore a valuable tool to interpret the model. 
For $s_{i,j,p}^{0\rightarrow1}(\mathscr{H}_t) = s_{i,j,GCP}^{0\rightarrow1}(\mathscr{H}_t)$, the first common partner has the multiplicative effect $2^{\alpha_p^{0\rightarrow 1}}$, while the 20th common partner has the multiplicative  effect $(21/20)^{\alpha_p^{0\rightarrow1}} = 1.05^{\alpha_p^{0\rightarrow1}}$ on the incidence intensity between actors $i$ and $j$.
This illustrates how dependence between counting processes is induced via a shared history $\mathscr{H}_t$: 
assuming that actors $i$ and $j$ are in an interaction at time $t$ and $N_{j,h}^{0\rightarrow 1}(t) >0$, a jump in $N_{i,h}^{0\rightarrow 1}(t)$ affects the intensity of a jump in $N_{i,j}^{1\rightarrow 0}(t)$ if the statistic $s_{i,j,GCP}^{1\rightarrow0}(\mathscr{H}_t)$ is included in the model.


\section{Scalable Block-Coordinate Ascent Algorithm}
\label{sec:computing}

Given the paths of $N_{i,j}^{\,0\rightarrow 1}(t)$ and $N_{i,j}^{\,1\rightarrow 0}(t)$ with $(i,j) \in \dyadset$ and $t \in \timeset$, we propose a scalable method to  
 estimate $\bm{\theta}^ {0\rightarrow 1}$ and $\bm{\theta}^ {\,1\rightarrow 0}$ maximizing the log-likelihood
\be
\label{eq:llh_joint}
\ell^{\,\star}(\bm{\theta}^{0\rightarrow 1},\bm{\theta}^ {\,1\rightarrow 0}) &\propto& \ell^{\,0\rightarrow 1}(\bm{\theta}^{0\rightarrow 1}) + \ell^ {\,1\rightarrow 0}(\bm{\theta}^ {\,1\rightarrow 0}),
\ee
which is separable with respect to $\bm{\theta}^ {0\rightarrow 1}$ and $\bm{\theta}^ {\,1\rightarrow 0}$.
This separability stems from the definition of the DEM as a multivariate counting process where the formation and dissolution intensities are conditionally independent given the sufficient statistics, time, and current interaction status of dyads.
Thus, we independently estimate the incidence and duration model using the same procedure. 
Redefining $\bm{\alpha} \in \mathbb{R}^P$, let $\bm{\theta} = (\bm{\alpha}, \bm{\beta},\bm{\gamma})$, $\actorsubset(t)$, $\ell(\bm{\theta})$, and $\bm{s}_{i,j}(\mathscr{H}_t)$ correspond to either $\bm{\theta}^{0 \rightarrow 1}$ or $\bm{\theta}^{1 \rightarrow 0}$, $\ell^{\,0\rightarrow 1}(\bm{\theta}^{\,0\rightarrow 1})$ or $\ell^{\,1\rightarrow 0}(\bm{\theta}^{\,1\rightarrow 0})$, $\actorsubset^{0 \rightarrow 1}(t)$ or $\actorsubset^{1 \rightarrow 0}(t)$, and $\bm{s}_{i,j}^{\,0\rightarrow 1}(\mathscr{H}_t)$ or $\bm{s}_{i,j}^{\,1\rightarrow 0}(\mathscr{H}_t)$, depending on estimating $\bm{\theta}^{0 \rightarrow 1}$ or $\bm{\theta}^{1 \rightarrow 0}$. 
The log-likelihood of each sub-model in \eqref{eq:llh_joint} is then:
\be
\label{eq:llh_0_1}
\ell(\bm{\theta}) &\propto& \dsum_{t\,\in\, \baselinechangeset\,\cup\, \eventsoccurset }\, \dsum_{(i,j) \,\in\, \actorsubset(t)}\,y_{i,j,t}\log \lambda_{i,j}\left(t \mid\mathscr{H}_t, \bm{\theta}\right) - \int_{t^\star}^t \lambda_{i,j}(u \mid \mathscr{H}_u, \bm{\theta}) \, \mathrm{d}u,
\ee
where $y_{i,j,t} \in \{0,1\}$ indicates whether a formation or dissolution event occurred between actors $i$ and $j$ in the time interval from $t^\star$ to $t$.
For the chronologically first time point in this set, we set $t^\star = 0$.
The estimate of $\bm{\theta}$ in the $k$th iteration of the algorithm is $\bm{\theta}^{(k)}$. 

The Newton-Raphson method is the state-of-the-art technique for estimating $\bm{\theta}$ in REMs \citep{butts_relational_2008, stadtfeld_dynamic_2017}, updating $\bm{\theta}^{(k+1)}$ by the following rule:  
\be
\label{eq:nr}
\bm{\theta}^{(k+1)} &=& \bm{\theta}^{(k)} -  \bm{\Sigma}(\bm{\theta}^{(k)})^{-1}\, \bm{g}(\bm{\theta}^{(k)}), 
\ee
where $\bm{g}(\bm{\theta}^{(k)}) \coloneqq \nabla_{\bm{\theta}} \,\ell(\bm{\theta}) \big|_{\bm{\theta} =\bm{\theta}^{(k)}}$ and $ \bm{\Sigma}(\bm{\theta}^{(k)}) \coloneqq \nabla_{\bm{\theta}}^2\, \ell(\bm{\theta}) \big|_{\bm{\theta} = \bm{\theta}^{(k)}}$ denote the gradient and Hessian of \eqref{eq:llh_0_1} evaluated at $\bm{\theta}^{(k)} \in \mathbb{R}^{P+N+Q}$, respectively.
\hide{
We partition $\Sigma(\bm{\theta})$ along $\bm{\alpha}, \bm{\beta}$, and $\bm{\gamma}$:
\be 
\bm{\Sigma}(\bm{\theta}) &=& \begin{pmatrix}
\bm{\Sigma}_{\bm{\alpha},\bm{\alpha}}  & \bm{\Sigma}_{\bm{\alpha},\bm{\beta}} & \bm{\Sigma}_{\bm{\alpha},\bm{\gamma}}\\
\bm{\Sigma}_{\bm{\alpha},\bm{\beta}}^\top  & \bm{\Sigma}_{\bm{\beta}, \bm{\beta}}   & \bm{\Sigma}_{\bm{\beta}, \bm{\gamma}}\\
\bm{\Sigma}_{\bm{\alpha},\bm{\gamma}}^\top  & \bm{\Sigma}_{\bm{\beta}, \bm{\gamma}}^\top & \bm{\Sigma}_{\bm{\gamma}, \bm{\gamma}},
\label{eq:hessian}
\end{pmatrix}
\ee
meaning that, e.g., matrix $\bm{\Sigma}_{\bm{\alpha},\bm{\beta}}$ is the block of the Hessian matrix concerning $\bm{\alpha}$ and $\bm{\beta}$. 
}
Assuming fixed $P= O(1)$ and $N\geq 2$, the per-iteration computational complexity for this algorithm is $O(N^2\,( N + Q)^2 \, (Q + M))$. 
In most real-world settings, data is available across many actors (large $N$) and over an extensive time interval (large $Q$), making the estimation of $\bm{\theta}$ via \eqref{eq:nr} impractical. 
To bypass this computational burden, we devise a block-coordinate ascent algorithm: 
\bi
\item[] {\bf Step 1:}
Set $\bm{\alpha}^{(k+1)}$ such that $\ell( \bm{\alpha}^{(k+1)}, \bm{\beta}^{(k)}, \bm{\gamma}^{\,(k)}) \,\geq\, \ell( \bm{\alpha}^{(k)}, \bm{\beta}^{(k)}, \bm{\gamma}^{\,(k)})$ by a Newton-Raphson update.
\item[] {\bf Step 2:}
Set $ \bm{\beta}^{(k+1)}$  such that $\ell( \bm{\alpha}^{(k+1)}, \bm{\beta}^{(k+1)}, \bm{\gamma}^{\,(k)}) \,\geq\, \ell( \bm{\alpha}^{(k+1)}, \bm{\beta}^{(k)}, \bm{\gamma}^{\,(k)})$ by a Minorization-Maximization update \eqref{eq:update_MM}.
\item[] {\bf Step 3:}
Set $ \bm{\gamma}^{\,(k+1)}$ such that $\ell( \bm{\alpha}^{(k+1)}, \bm{\beta}^{(k+1)}, \bm{\gamma}^{\,(k+1)}) \,\geq\, \ell( \bm{\alpha}^{(k+1)}, \bm{\beta}^{(k+1)}, \bm{\gamma}^{\,(k)})$ by a closed-form update \eqref{eq:update_gamma}.
\ei
\hide{

The following proposition details the computational advantages of this blockwise algorithm. 

\begin{proposition}
\label{prop_prop}
{\em
 Let $l(\theta)$ be the log-likelihood function defined in (6), and let $N$, $M$, and $Q$ denote the number of actors, observed events, and baseline change points, respectively. The proposed Block-Coordinate Ascent algorithm satisfies the following properties:
 \begin{itemize}
     \item[(a)] The complexity per iteration  $O(N^2 + M \times N + M)$. 
     \item[(b)] we obtain global convergence the maximum likelihood estimator $\bm{\theta}^{\star}$, i.e., $\bm{\theta}^{(k)}\rightarrow \bm{\theta}^{\star}$ with $k \rightarrow \infty$ for any starting value $\bm{\theta}^{(0)}$. 
     \item[(c)] Our algorithm enjoys the ascent property, $\ell( \bm{\alpha}^{(k+1)}, \bm{\beta}^{(k+1)}, \bm{\gamma}^{\,(k+1)}) \,\geq\, \ell( \bm{\alpha}^{(k)}, \bm{\beta}^{(k)}, \bm{\gamma}^{\,(k)})$, making the algorithm robust and reliable.
 \end{itemize}

}
\end{proposition}
Proposition \ref{prop_prop} if proved in the Supplement. 
compared to the Newton-Raphson update in \eqref{eq:nr} is reduced from $O((P+ N + Q)^3)$ to
Second, there is no need to store large matrices at any step, reducing memory usage.
Further, note that this algorithm naturally extends to standard REMs. 
 We declare convergence once both $\spectralnormvec{\bm{\theta}^{(k+1)} - \bm{\theta}^{(k)}}$ and $|(\ell(\bm{\theta}^{(k+1)}) - \ell(\bm{\theta}^{(k)}))\, /\, \ell(\bm{\theta}^{(k)})|$ are below a given threshold, typically set at $10^{-3}$. 
We describe the algorithm in the ensuing paragraphs. 

}
The blockwise algorithm offers several computational advantages:
first,
the complexity per iteration compared to the Newton-Raphson update in \eqref{eq:nr} is reduced from $O((N + Q)^2\,N^2\, (Q + M) )$ to $O(N^2 + Q + MN)$. 
This improvement is achieved by restricting matrix inversions to the first step, leveraging the low dimensionality of  $\bm{\alpha}$  to ensure scalability with respect to $N$ and $Q$. 
In that step, we also employ caching algorithms.
Second, there is no need to store large matrices at any step, reducing memory usage.
Third, 
we obtain global convergence to the maximum likelihood estimator $\bm{\theta}^{\star}$, i.e., $\bm{\theta}^{(k)}\rightarrow \bm{\theta}^{\star}$ with $k \rightarrow \infty$ for any starting value $\bm{\theta}^{(0)}$. 
This property holds since the likelihood function \eqref{eq:llh_0_1} can be decomposed into sums of likelihoods of Poisson-distributed random variables \citep{fritz_all_2023}, making $\ell(\bm{\theta})$ a strictly concave function. 
We therefore seed our algorithm by setting $\bm{\theta}^{(0)} = \bm{0}_{P+N+Q}$,  
being a vector filled with zeros.
Fourth, contrasting the Newton-Raphson method, our algorithm enjoys the ascent property, $\ell( \bm{\alpha}^{(k+1)}, \bm{\beta}^{(k+1)}, \bm{\gamma}^{\,(k+1)}) \,\geq\, \ell( \bm{\alpha}^{(k)}, \bm{\beta}^{(k)}, \bm{\gamma}^{\,(k)})$, making the algorithm robust and reliable.
Fifth, our algorithm naturally extends to standard REMs. 
 Convergence is declared once both $\spectralnormvec{\bm{\theta}^{(k+1)} - \bm{\theta}^{(k)}}$ and $|(\ell(\bm{\theta}^{(k+1)}) - \ell(\bm{\theta}^{(k)}))\, /\, \ell(\bm{\theta}^{(k)})|$ are below a given threshold, typically set at $10^{-3}$. 
We next describe the algorithm in detail. 

\paragraph*{Step 1: Update of $\bm{\alpha}$.}
Similar to \eqref{eq:nr}, we employ Newton-Raphson updates for $\bm{\alpha}$ with:
\beno
\label{eq:step1}
\nabla_{\bm{\alpha}}\, \ell(\bm{\alpha} , \bm{\beta}^{(k)}, \bm{\gamma}^{\,(k)})\big|_{\bm{\alpha} = \bm{\alpha}^{(k)}} &=\,\dsum_{t\,\in\, \eventsoccurset}  \, \dsum_{(i,j) \,\in\, \actorsubset(t)}\bm{s}_{i,j}(\mathscr{H}_t) \left(y_{i,j,t} - \dint_{t^\star}^t \lambda_{i,j}(u\mid \mathscr{H}_u, \bm{\theta}^{(k)})  \,\text{d}u\right) \s\\
\nabla_{\bm{\alpha}}^2\, \ell(\bm{\alpha} , \bm{\beta}^{(k)}, \bm{\gamma}^{\,(k)})\big|_{\bm{\alpha} = \bm{\alpha}^{(k)}} &=\, - \dsum_{t\,\in\, \eventsoccurset}\, \dsum_{(i,j) \,\in\, \actorsubset(t)} \bm{s}_{i,j}(\mathscr{H}_t) \, \bm{s}_{i,j}(\mathscr{H}_t)^\top \left(\dint_{t^\star}^t \lambda_{i,j}(u\mid \mathscr{H}_u, \bm{\theta}^{(k)}) \, \text{d}u\right).
\ee
Because the intensity function is piecewise-constant, the integral $\int_{t^\star}^t \lambda_{i,j}(u \mid \mathscr{H}_u, \bm{\theta}) \, \text{d}u$ admits an exact computation. 
Inspired by caching algorithms introduced in \citet{vu_continuous-time_2011}, we only update the summary statistics corresponding to pairs affected by $d$ after observing the event $d$ at time $t$. 
Because the summary statistics in Section \ref{sec:summary_statistics} are defined locally, we assume that each interaction solely affects at most $O(N)$ pairs. 
Under $P = O(1)$, the complexity of this step is then $O(N^2 + MN)$, eliminating the dependence on the number of change points of the baseline intensity.

\paragraph*{Step 2: Update of $\bm{\beta}$.}
A Newton-Raphson update for $\bm{\beta}$ is computationally infeasible due to the dimension of $\bm{\beta}$ being $N$. 
To avoid this bottleneck,
we derive a Minorization-Maximization (MM) algorithm \citep{langeOptimizationTransferUsing2000}. 
Within this algorithm, we define a surrogate function with a closed-form solution whose maximizer guarantees $\ell( \bm{\alpha}^{(k+1)}, \bm{\beta}^{(k+1)}, \bm{\gamma}^{\,(k)}) \,\geq\, \ell( \bm{\alpha}^{(k+1)}, \bm{\beta}^{(k)}, \bm{\gamma}^{\,(k)})$. 
To derive this function, we first reformulate $\ell(\bm{\alpha}^{(k+1)}, \bm{\beta}, \bm{\gamma}^{\,(k)})$, being a function of $\bm{\beta}$ with fixed $\bm{\alpha} =\bm{\alpha}^{(k+1)}$ and $\bm{\gamma} =\bm{\gamma}^{(k)}$:
\be
\label{eq:step3}
\ell(\bm{\alpha}^{(k+1)}, \bm{\beta}, \bm{\gamma}^{\,(k)} ) &\propto& \dsum_{(i,j) \,\in\, \dyadset}  (\log b_i + \log b_j)\left(\dsum_{t\,\in\, \eventsoccurset} y_{i,j,t}\right)  -  b_i\,b_j\, \left(\dsum_{t\,\in\, \eventsoccurset} b_{i,j,t}\right),
\ee
where
\beno
b_{i,j,t} \,\coloneqq\, \left(\,\dint_{t^\star}^t  \exp\left(f(u, \bm{\gamma}^{(k)})\right)\, \text{d}u  \right)\,\exp\left(\bm{\alpha}^{(k+1)}\, \bm{s}_{i,j}(\mathscr{H}_t)\right) 
\ee
and $b_i \,\coloneqq\,\exp(\beta_i)$ for $t\,\in\, \eventsoccurset$ and $(i,j)\in \dyadset$.
By the AM–GM inequality, we get 
\be
\label{eq:step4}
b_{i}\, b_{j}  &\leq& \dfrac{b_{j}^{(k)}}{2\, b_{i}^{(k)}}\, b_{i}^2 + \dfrac{b_{i}^{(k)}}{2\, b_{j}^{(k)}}\, b_{j}^2,
\ee
allowing us to define a surrogate function $m(\bm{\beta} \mid \bm{\alpha}^{(k+1)}, \bm{\beta}^{(k)}, \bm{\gamma}^{\,(k)})$: 
\beno
\label{eq:minorizer}
\ell(\bm{\alpha}^{(k+1)}, \bm{\beta}, \bm{\gamma}^{\,(k)} ) &\geq& \dsum_{(i,j) \,\in\, \dyadset}  (\log b_i + \log b_j)\left(\dsum_{t\,\in\, \eventsoccurset} y_{i,j,t}\right)  \s\\ 
&-&  \left(\dfrac{b_{j}^{(k)}}{2\, b_{i}^{(k)}} b_{i}^2 + \dfrac{b_{i}^{(k)}}{2\, b_{j}^{(k)}} b_{j}^2\right)\, \left(\dsum_{t\,\in\, \eventsoccurset} b_{i,j,t}\right)\s \\
&\eqqcolon&m(\bm{\beta} \mid \bm{\alpha}^{(k+1)}, \bm{\beta}^{(k)}, \bm{\gamma}^{\,(k)}).
\ee
Since equality holds in \eqref{eq:step4} with $b_i= b_i^{(k)}$ and $b^{(k)}_i\,\coloneqq\,\exp(\beta_i^{(k)})$ for all $i \in \actorset$, 
we get:
\be
m(\bm{\beta} \mid \bm{\alpha}^{(k+1)}, \bm{\beta}^{(k)}, \bm{\gamma}^{\,(k)}) &\leq& \ell(\bm{\alpha}^{(k+1)}, \bm{\beta} , \bm{\gamma}^{\,(k)}) &\text{for all }\bm{\beta}\in\mathbb{R}^N \\
m(\bm{\beta}^{(k)} \mid \bm{\alpha}^{(k+1)}, \bm{\beta}^{(k)}, \bm{\gamma}^{\,(k)}) &=& \ell(\bm{\alpha}^{(k+1)}, \bm{\beta}^{(k)}, \bm{\gamma}^{\,(k)}),
\label{eq:properties}
\ee
making
 $m(\bm{\beta} \mid \bm{\alpha}^{(k+1)}, \bm{\beta}^{(k)}, \bm{\gamma}^{\,(k)})$ a minorizer of $\ell(\bm{\alpha}^{(k+1)},\bm{\beta}, \bm{\gamma}^{\,(k)})$ at $\bm{\beta}^{(k)}$.
 Applying these inequalities shows that setting $ \bm{\beta}^{(k+1)}$ to the value maximizing $m(\bm{\beta} \mid \bm{\alpha}^{(k+1)}, \bm{\beta}^{(k)}, \bm{\gamma}^{\,(k)})$ with respect to $\bm{\beta}$, guarantees $\ell( \bm{\alpha}^{(k+1)}, \bm{\beta}^{(k+1)}, \bm{\gamma}^{\,(k)}) \,\geq\, \ell( \bm{\alpha}^{(k+1)}, \bm{\beta}^{(k)}, \bm{\gamma}^{\,(k)})$. 
In contrast to $\ell(\bm{\alpha}^{(k+1)}, \bm{\beta}, \bm{\gamma}^{\,(k)})$, $m(\bm{\beta} \mid \bm{\alpha}^{(k+1)}, \bm{\beta}^{(k)}, \bm{\gamma}^{\,(k)})$ has a closed form solution for $\bm{\beta}$ for $i \in \actorset$:
\be
\label{eq:update_MM}
\beta_i^{(k+1)} &=& \log\left(\sqrt{b_i^{(k)} \dfrac{\dsum_{j\,\neq\, i}\,\dsum_{t\,\in\, \eventsoccurset} y_{i,j,t}}{ \dsum_{j \,\neq\, i} \, \dsum_{t\,\in\, \eventsoccurset} b_{i,j,t}\, b_j^{(k)}}}\,\right).
\ee
These cyclical updates for $i = 1, \ldots, N$ have the algorithmic complexity $O(N\,(N + M))$.

\paragraph*{Step 3: Update of $\bm{\gamma}$.}
With $h_{i,j,t} \,\coloneqq\, \int_{t^\star}^t \exp(\bm{\alpha}^{(k+1)}\, \bm{s}_{i,j}(\mathscr{H}_u) + {\beta}_i^{(k+1)} + {\beta}_j^{(k+1)})\,\text{d}u$ and $h_t \,\coloneqq\,\exp(f(t, \bm{\gamma}))$ for $t\,\in\, \baselinechangeset \cup \eventsoccurset$ and $(i,j)\in \dyadset$,  
we can sort the summands of $\ell(\bm{\alpha}^{(k+1)}, \bm{\beta}^{(k+1)}, \bm{\gamma})$ according to the intervals defined through the timepoints where the baseline intensity changes,  $0 = c_0 < c_1< \ldots < c_Q$:
\be
\label{eq:llh_gamma}
\ell(\bm{\alpha}^{(k+1)}, \bm{\beta}^{(k+1)}, \bm{\gamma}) &\propto& \dsum_{(i,j) \,\in\, \dyadset}\, \dsum_{q\, =\, 1}^Q\, \dsum_{c_{q-1} \,\leq \,t\, < \,c_q}  y_{i,j,t}\, \log h_t  -  h_t\,  h_{i,j,t}.
\ee 
This function is separable with respect to all coordinates of $\bm{\gamma}$ with a closed-form solution:
\be
\label{eq:update_gamma}
\gamma^{\,(k+1)}_q &=& \log \left(\dfrac{\dsum_{(i,j) \,\in\, \dyadset}\, \dsum_{c_{q-1} \,\leq \,t\, < \,c_q} y_{i,j,t}}{\dsum_{(i,j) \,\in\, \dyadset}\, \dsum_{c_{q-1} \,\leq \,t\, < \,c_q}  h_{i,j,t}}\right).
\ee
Employing update \eqref{eq:update_gamma} for $q = 2, \dots, Q$ has the algorithmic complexity of $O(N^2 + Q+ MN)$.

\hide{
This function is separable with respect to all coordinates of $\bm{\gamma}$ with a closed-form solution: \be \label{eq:update_gamma} \gamma^{,(k+1)}q &=& \log \left(\dfrac{\dsum{(i,j) ,\in, \dyadset}, \dsum_{c_{q-1} ,\leq ,t, < ,c_q} y_{i,j,t}}{\dsum_{(i,j) ,\in, \dyadset}, \dsum_{c_{q-1} ,\leq ,t, < ,c_q} h_{i,j,t}}\right). \ee A naive evaluation of \eqref{eq:update_gamma} would evaluate the risk set for each time slice independently, yielding a multiplicative complexity of $O(Q(N^2 + M))$. However, by maintaining an asynchronous timeline where dyadic risk intervals are splintered only when a dyad's specific sufficient statistics change, the total number of piecewise-constant intervals across the entire network is bounded by $O(N^2 + M \cdot N)$. Furthermore, by analytically integrating the exposure times across the $Q$ baseline intervals in a single pass, the updates for all $q = 2, \dots, Q$ are computed simultaneously. This reduces the algorithmic complexity of updating $\bm{\gamma}^{(k+1)}$ to strictly $O(N^2 + M \cdot N + Q)$.
}

\paragraph*{Uncertainty Quantification.}
Inference primarily focuses on quantifying the uncertainty of the parameters $\hat{\bm{\alpha}}$, treating $\bm{\beta}$ and $\bm{\gamma}$ as nuisance parameters.
Isolating this variance requires extracting the block corresponding to $\hat{\bm{\alpha}}$ of the inverse Fisher info, $(-\bm{\Sigma}(\hat{\bm{\theta}}))^{-1}$, evaluated at the converged estimates $\hat{\bm{\theta}}$.  
By block inversion, this sub-matrix, denoted by $\bm{\Lambda}(\hat{\bm{\theta}})$, is given by:
\be
\label{eq:inv_fisher}
\bm{\Lambda}(\hat{\bm{\theta}}) &=& \left(\bm{X}_{\hat{\bm{\alpha}}, \hat{\bm{\beta}},\hat{\bm{\gamma}}}\, \bm{Y}_{\hat{\bm{\beta}},\hat{\bm{\gamma}}}^{-1}\, \bm{X}_{\hat{\bm{\alpha}}, \hat{\bm{\beta}},\hat{\bm{\gamma}}}^\top - \bm{\Sigma}_{\hat{\bm{\alpha}},\hat{\bm{\alpha}}} \right)^{-1},
\ee
with 
\beno 
\bm{X}_{\hat{\bm{\alpha}}, \hat{\bm{\beta}},\hat{\bm{\gamma}}} &=& \left(\bm{\Sigma}_{\hat{\bm{\alpha}},\hat{\bm{\beta}}} ~~ \bm{\Sigma}_{\hat{\bm{\alpha}},\hat{\bm{\gamma}}}\right) \text{ and  } \bm{Y}_{\hat{\bm{\beta}},\hat{\bm{\gamma}}} &=& \begin{pmatrix}
\bm{\Sigma}_{\hat{\bm{\beta}}, \hat{\bm{\beta}}} & \bm{\Sigma}_{\hat{\bm{\beta}}, \hat{\bm{\gamma}}} \\
\bm{\Sigma}_{\hat{\bm{\beta}}, \hat{\bm{\gamma}}}^\top & \bm{\Sigma}_{\hat{\bm{\gamma}}, \hat{\bm{\gamma}}} \\
\end{pmatrix},
\ee
where, e.g., matrix $\bm{\Sigma}_{\hat{\bm{\alpha}},\hat{\bm{\beta}}}$ is the sub-block of $\bm{\Sigma}$ associated with $\hat{\bm{\alpha}}$ and $\hat{\bm{\beta}}$. 

Establishing the asymptotic normality of $\hat{\bm{\alpha}}$ requires addressing the temporal dependence inherent in counting processes and clarifying the inferential framework. 
Two regimes are relevant in this setting: observing a fixed set of actors over an increasing time horizon, $T\to \infty$ with fixed $N$, or observing a growing population over a fixed time horizon, $N\to \infty$ with fixed $T$ \citep{fritz2026a}. 
Either setting introduces high-dimensional challenges, since the dimension of $\bm{\beta}$ or $\bm{\gamma}$ may increase with $M$. 
Semiparametric profile likelihood theory \citep{murphy2000} and martingale central limit theorems for counting processes \citep{andersen_coxs_1982} provide a rigorous foundation for inference in this setting but require that the true parameter lies in a compact space. 
This condition precludes that any individual actor remains isolated, since the observed information will be singular under $\beta_i \to -\infty$. 
The profile likelihood framework ensures that regular asymptotic normality holds provided the nuisance dimension satisfies $(N + Q)^2 / M \to 0$ as $M \to \infty$. 
For $Q\to M$, \eqref{eq:llh_0_1} converges to the partial likelihood where the normal approximation from \citet{perry_point_2013} applies.
Evidence from Simulation Study 5 (Supplement A.3.3) indicates consistency and normality still hold when $Q$ is misspecified.


\hide{
\alert{Note that results from Beta Models could be adapted to further approximate $Y_{\bm{\alpha}, \bm{\beta},\bm{\gamma}}^{-1}$ by a diagonal matrix if needed.}
}

\section{Simulation Study}
\label{sec:simulation_study}


We evaluate our estimator's performance across three simulation studies targeting parameter recovery, uncertainty quantification, and computational improvement.
The experimental design involves $S=\mbox{1,000}$ simulated datasets for several actor population sizes $N$ with covariates:
one continuous covariate, $\bm x = (x_{1}, \ldots, x_{N})$ with $X_i \overset{\text{i.i.d}}{\sim} \mathcal{N}(0,1)$, and one discrete covariate, $\bm w = (w_{1}, \ldots, w_{N})$ with $W_i \overset{\text{i.i.d}}{\sim} \text{Categorical}(1/3, 1/3, 1/3)$, is available.
These covariates inform the summary statistics vectors $\bm{s}_{i,j}^{0 \rightarrow 1} = (s_{i,j,CCP}, s_{i,j,\bm{x}}, s_{i,j,\bm{w}})^\top$ and $\bm{s}_{i,j}^{1 \rightarrow 0} = (s_{i,j,NI}, s_{i,j,\bm{x}}, s_{i,j,\bm{w}})^\top$, with their dependence on the history $\mathscr{H}_t$ suppressed for notational clarity. 
The corresponding true parameter vectors are fixed at $\bm{\alpha}^{\,0\rightarrow 1} = (-1/2,1,1/2)$ and $\bm{\alpha}^{\,1\rightarrow 0} = (1/2,1/2,1/2)$, 
while the popularity parameters $\bm{\beta}$ are Gaussian-distributed, $\bm{\beta}^{0\rightarrow 1}$ with $\beta_i^{0\rightarrow 1} \overset{\text{i.i.d}}{\sim} \mathcal{N}(- 6- 1/10\,\log(N),1)$ 
and $\bm{\beta}^{1\rightarrow 0}$ with $\beta_i^{1\rightarrow 0} \overset{\text{i.i.d}}{\sim} \mathcal{N}(8/5 - 1/10\,\log(N),1)$,
decreasing proportional to $-\log(N)$ to induce sparsity.  
 We specify a baseline by $\bm{\gamma} = \bm{\gamma}^{0\rightarrow 1} = \bm{\gamma}^{\,1\rightarrow 0}$ which decays linearly from 0 to -0.1 over $\mathscr{T} = [0, 10^4]$.
Supplement A provides details on the simulation algorithm, evaluation measures, and further results.

\paragraph*{Simulation Study 1: Parameter Estimation.}
In the first simulation study, we assess how accurately our novel algorithm recovers the correct model for $N = 500$ actors. 
Table \ref{tbl:simulation} demonstrates that $\bm{\alpha}$ is estimated with high accuracy. 
The empirical coverage probability matches the nominal $.95$ level, indicating reliable uncertainty quantification.
For the current common partner statistic, we note a higher RMSE compared to the other statistics, which may be due to the statistic using less data by only relying on the current status of the network rather than its entire history. 
Supplement A.3.1 provides results  on model selection using the Akaike Information Criterion and the normality of estimators. 


\hide{
\begin{table}[!t]
\caption{\alert{Still need to update}Simulation Study 1:
      For each effect, we report the Bias ($\hat{\bm{\alpha}}-\bm{\alpha}^\star$), RMSE (Root-Mean-Squared Error), and CP (Coverage Probability). The dependence of the sufficient statistics on the history $\mathscr{H}_t$ suppressed for notational clarity. \label{tbl:simulation}} 
\begin{center}
\begin{tabular}{lrrrclrrr}
\hline
\multicolumn{4}{c}{\bfseries Incidence ($\alpha^{0\rightarrow 1}$)}&\multicolumn{1}{c}{\bfseries }&\multicolumn{4}{c}{\bfseries Duration ($\alpha^{\,1\rightarrow 0}$)}\tabularnewline
\multicolumn{1}{l}{\bfseries Statistic}&\multicolumn{1}{c}{\bfseries Bias}&\multicolumn{1}{c}{\bfseries RMSE}&\multicolumn{1}{c}{\bfseries CP}&\multicolumn{1}{c}{}&\multicolumn{1}{l}{\bfseries Statistic}&\multicolumn{1}{c}{\bfseries Bias}&\multicolumn{1}{c}{\bfseries RMSE}&\multicolumn{1}{c}{\bfseries CP}\tabularnewline
\hline
$s_{i,j,CCP}$&$-.010$&.074&.951&&$s_{i,j,NI}$&.000&.010&.945\tabularnewline
$s_{i,j,\bm{x}}$&$-.002$&.010&.944&&$s_{i,j,\bm{x}}$&$.002$&.011&.939\tabularnewline
$s_{i,j,\bm{y}}$&$-.001$&.010&.957&&$s_{i,j,\bm{y}}$&$.001$&.011&.944\tabularnewline
\hline
\end{tabular}\end{center}
\end{table}
}

\begin{table}[!tbp]
\caption{Simulation Study 1:
      For each effect, we report the Bias ($\hat{\bm{\alpha}}-\bm{\alpha}^\star$), RMSE (Root-Mean-Squared Error), and CP (Coverage Probability). The dependence of the sufficient statistics on the history $\mathscr{H}_t$ is suppressed for notational clarity. \label{tbl:simulation}} 
\begin{center}
\begin{tabular}{lrrrclrrr}
\hline
\multicolumn{1}{l}{\bfseries }&\multicolumn{3}{c}{\bfseries Incidence ($\alpha^{0\rightarrow 1}$)}&\multicolumn{1}{l}{\bfseries }&\multicolumn{4}{c}{\bfseries Duration ($\alpha^{1\rightarrow 0}$)}\tabularnewline
\multicolumn{1}{l}{}&\multicolumn{1}{r}{Bias}&\multicolumn{1}{r}{RMSE}&\multicolumn{1}{r}{CP}&\multicolumn{1}{c}{}&\multicolumn{1}{c}{}&\multicolumn{1}{r}{Bias}&\multicolumn{1}{r}{RMSE}&\multicolumn{1}{r}{CP}\tabularnewline
\hline
$s_{i,j,CCP}$&$-$.009&.074&.946&&$s_{i,j,NI}$&.000&.010&.956\tabularnewline
$s_{i,j,\bm{x}}$&.001&.010&.961&&$s_{i,j,\bm{x}}$&.000&.011&.949\tabularnewline
$s_{i,j,\bm{w}}$&.000&.010&.950&&$s_{i,j,\bm{w}}$&.000&.011&.962\tabularnewline
\hline
\end{tabular}\end{center}
\end{table}

\paragraph*{Simulation Study 2: Estimation Error under Increasing Actors.}
We rely on an analogous setup as in Simulation Study 1, with the only difference being that the number of actors varies from $50$ to $\mbox{1,000}$. 
We assess the estimation error of all parameters. 

\begin{figure}[t!]
  \centering
  \includegraphics[width=0.75\textwidth]{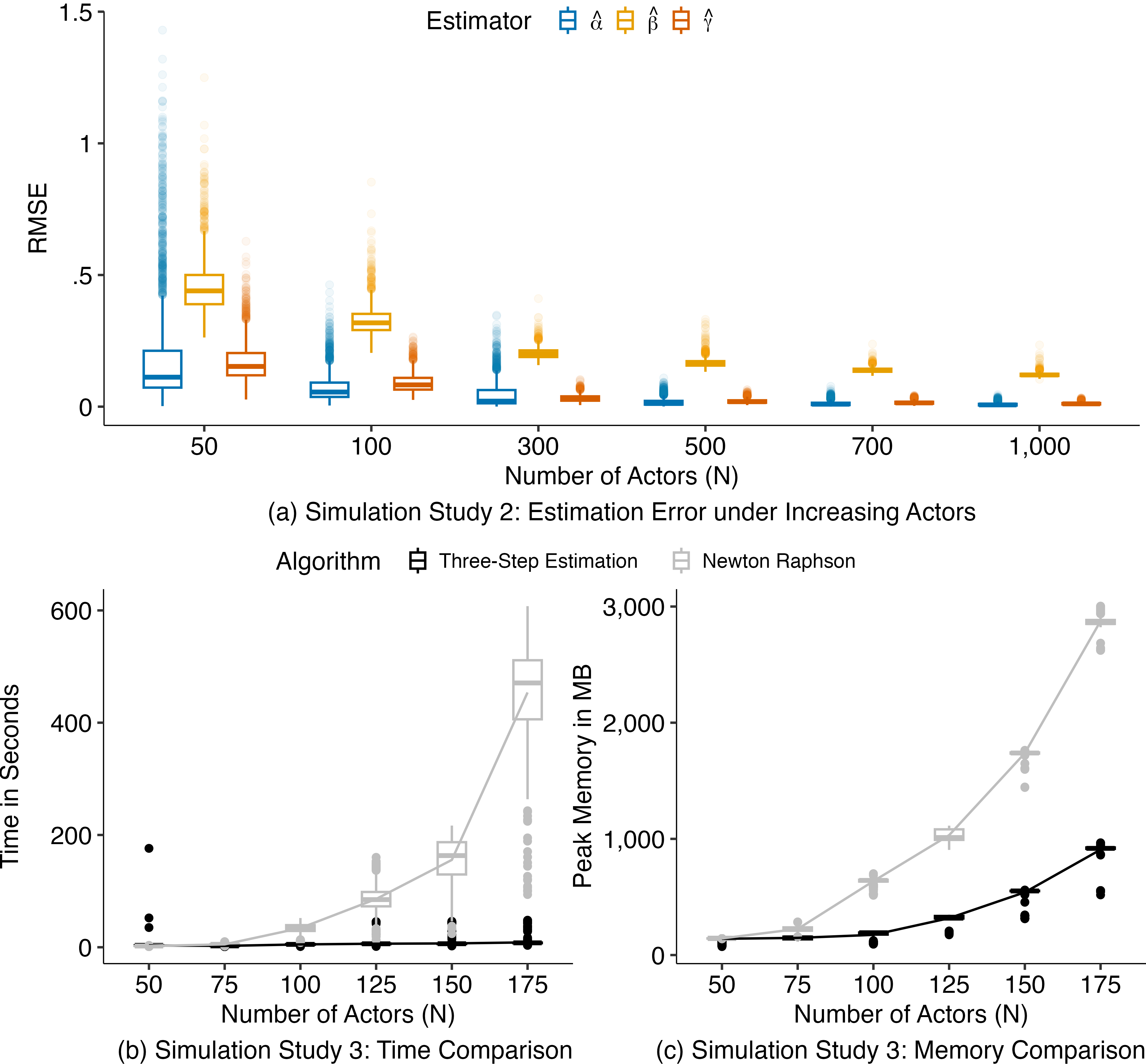}  
    \caption{Simulation Study 2: (a) RMSE of $\hat{\bm{\alpha}}$ (blue), $\hat{\bm{\beta}}$ (yellow), and $\hat{\bm{\gamma}}$ (orange) pooled over incidence and duration model. Simulation Study 3: Comparison of computational time (b) and memory needed for estimation  (c) for different number of actors of our proposed three-step estimator and the state-of-the-art Newton-Raphson method.}
  \label{fig:simulation}
\end{figure}

  For both sub-models, Figure \ref{fig:simulation} (a) shows that the RMSE of all estimated parameters decreases as the number of actors $N$ increases. 
  The baseline intensity and summary statistics coefficients exhibit lower estimation error than the popularity parameters, whose dimensionality grows with $N$.
  This behavior is expected due to the increasing dimension of the popularity effects. 
  These results provide empirical evidence for the consistency of our estimators in regimes where the number of parameters is a function of the number of actors.


\paragraph*{Simulation Study 3: Computational Improvement.}
Finally, we compare the computational efficiency of our novel algorithm (Section \ref{sec:computing}) to the Newton-Raphson approach. 
We measure computational speed by the execution time (in seconds) for estimating the model on a single dataset. 
Memory allocation is assessed by each algorithm's peak random access memory usage, to reflect the largest dataset a system can handle. 

Figure \ref{fig:simulation} (b) and (c) visualizes the execution time and memory usage for both algorithms under increasing number of actors.
The largest dataset we consider has only up to $175$ nodes, beyond which the Newton-Raphson algorithm becomes impractical. 
The results provide empirical evidence that our block-coordinate ascent algorithm improves upon state-of-the-art methods by several orders of magnitude.
From Figure \ref{fig:simulation} (c), it is evident that our proposed algorithm scales more efficiently than the Newton-Raphson method. 
Additionally, our approach exhibits lower variability in computing time and memory usage.
Both findings align with the theoretical complexity reduction derived in Section \ref{sec:computing}.

\section{Application to Physical and Digital Interactions}
\label{sec:applications}
We demonstrate the DEM in an application to the Copenhagen Networks Study \citep{sapiezynskiInteractionDataCopenhagen2019}.  
The study, conducted between 2012 and 2013, tracked 682 first-year students at the Technical University of Denmark over around 28 days. 
This data includes various forms of interactions, i.e., co-location (measured via Bluetooth), social media interactions (friendships on Facebook), and other communication channels (phone calls and text messages), which allows comparing the evolution of physical and digital interactions between the same actors over time. 
We can thus  identify how individuals prioritize different relationships and the circumstances under which they rely on digital or face-to-face interactions \citep{stopczynski2014}.
We study digital communication through call data, while co-location events act as a proxy for physical activities. 
By distinguishing when interactions begin and end, we can (i) fully utilize the available data and (ii) assess whether similar factors drive both the incidence and duration of events.
Several studies have used the Copenhagen Networks Study for descriptive analyses: 
\citet{sekaraFundamentalStructuresDynamic2016} demonstrate with it how interaction data with fine-grained temporal resolution can be used to examine group formation, while
\citet{EstimatingTieStrength2020} investigate the strength of interactions with the data. 
However, to our knowledge, no probabilistic models have been applied to this dataset. 



 Physical interactions are quantified by co-location events, measured via Bluetooth communication between mobile devices. 
 Bluetooth scans run every five minutes determine when people are close to one another. 
 We define a co-location event as the presence of two individuals in at least two consecutive scans. 
 Therefore, a physical interaction event $(i,j,t_b,t_e)$ indicates that student $i$ was in the same place as student $j$ from time $t_b$ to $t_e$.
 
Call data represent digital communication, where an event $(i,j,t_b,t_e)$ encodes a call between student $i$ and student $j$ between the time points $t_b$ and $t_e$. 
The sets $\actorsubset^{0 \rightarrow 1}(t)$ and $\actorsubset^{1 \rightarrow 0}(t)$ for $t\in \timeset$ are defined so that no student can be active in simultaneous calls (see Section \ref{sec:dem}).

Participants provided covariate information on their Facebook friends and gender. 
We included actors with this information who participated in at least one durational event in both communication modes (physical and digital), ensuring identifiable popularity coefficients.  
This amounts to $\mbox{155,316}$ physical and $\mbox{4,152}$ digital interactions among $N = 400$ actors.
In Supplementary B.6, we use instantaneous text messages from the participants to show the algorithm's applicability to relational event data.

\subsection{Model Specification} 
\label{sec:specification}

To examine endogenous effects, we use summary statistics from Section \ref{sec:summary_statistics}.  
We evaluate whether actors $i$ and $j$ currently share and have previously shared a common partner, and we account for pairwise past interactions. 
For call events, having a common partner at the time of the event is impossible, as an individual can participate in only one phone call at a time. 
Accordingly, we exclude the current common partner terms from the corresponding incidence and duration models.
We further incorporate exogenous covariates: Facebook friendship status ($z_{i,j,1} = 1$ if actors $i$ and $j$ are friends, and $0$ otherwise) and gender homophily ($z_{i,j,2} = \mathbb{I}(x_{i} = x_{j})$, where $x_{i} \in \{0,1\}$ indicates the gender of actor $i$).


Substantial heterogeneity in interaction frequency across actors for both physical and digital events (see Supplement B.2), motivates the inclusion of actor-specific popularity parameters in the model.
Interaction frequencies are expected to fluctuate considerably throughout the day since students are more likely to interact during the daytime hours. 
Thus, we let the baseline function  $f(t, \bm{\gamma})$ change hourly, implying $Q = 672$. 
We conduct a sensitivity analysis in Supplement B.5 showing that letting  $f(t, \bm{\gamma})$ change every two hours does not substantially affect the results.
While solving this problem is infeasible with state-of-the-art techniques, our novel algorithm enables model estimation in less than five minutes on a standard laptop.

\subsection{Results}
        
The estimates $\hat{\bm{\alpha}}$ reveal distinct patterns for incidence and duration of physical and digital interactions (popularity and temporal estimates are provided in Supplement B.3).


\hide{
\begin{table}[t!]
\caption{Parameter estimates $(\hat{\bm{\alpha}})$ and corresponding standard errors (SE) of the Durational Event Model applied to physical (co-location, columns 2-4) and digital (call, columns 5-7) interaction data. 
The column $2^{\hat{\bm{\alpha}}}$ shown for statistics  transformed by $\log(\,\cdot +1)$ represents the effect of the first change in the respective statistic. \label{tbl:results}} 
\begin{center}
\begin{tabular}{lrrrcrrr}
\hline
\multicolumn{1}{l}{\bfseries}&\multicolumn{3}{c}{\bfseries Physical (Co-location)}&\multicolumn{1}{l}{\bfseries }&\multicolumn{3}{c}{\bfseries Digital (Call)}\tabularnewline
\cline{2-4} \cline{6-8}
\multicolumn{1}{l}{\textbf{Statistic}}&\multicolumn{1}{c}{$\hat{\bm{\alpha}}$}&\multicolumn{1}{c}{ \bfseries SE}&\multicolumn{1}{c}{$2^{\hat{\bm{\alpha}}}$}&\multicolumn{1}{c}{}&\multicolumn{1}{c}{$\hat{\bm{\alpha}}$}&\multicolumn{1}{c}{\bfseries SE}&\multicolumn{1}{c}{$2^{\hat{\bm{\alpha}}}$}\tabularnewline
\hline
{ Incidence ($\hat{\bm{\alpha}}^{0\rightarrow 1}$)}&&&&&&&\tabularnewline
~~Current Common Partner&2.867&.006&7.295&&&&\tabularnewline
~~General Common Partner&.726&.007&1.654&&.224&.125&1.168\tabularnewline
~~Number Interaction&1.129&.005&2.187&&1.631&.039&3.097\tabularnewline
~~Friendship Match&.383&.010&&&5.687&.116&\tabularnewline
~~Both Female&$-$.021&.010&&&.203&.085&\tabularnewline
\hline
{ Duration ($\hat{\bm{\alpha}}^{\,1\rightarrow 0}$)}&&&&&&&\tabularnewline
~~Number Interaction&$-$.158&.005&.896&&$-$.274&.073&.827\tabularnewline
~~Current Interaction&$-$.102&.002&.932&&$-$.053&.024&.964\tabularnewline
~~Current Common Partner&$-$.312&.006&.806&&&&\tabularnewline
~~General Common Partner&.080&.009&1.057&&.518&.186&1.432\tabularnewline
~~Friendship Match&$-$.535&.010&&&$-$.337&.464&\tabularnewline
~~Both Female&$-$.018&.010&&&$-$.227&.320&\tabularnewline
\hline
\end{tabular}\end{center}
\end{table}
}

\begin{table}[t!]
\caption{Parameter estimates $(\hat{\bm{\alpha}})$ and corresponding standard errors (SE) of the Durational Event Model applied to physical (co-location, columns 2-4) and digital (call, columns 5-7) interaction data. 
The column $2^{\hat{\bm{\alpha}}}$ shown for statistics  transformed by $\log(\,\cdot +1)$ represents the effect of the first change in the respective statistic. \label{tbl:results}} 
\begin{center}
\begin{tabular}{lrrrcrrr}
\hline
\multicolumn{1}{l}{\bfseries }&\multicolumn{3}{c}{\bfseries Physical (Co-location)}&\multicolumn{1}{l}{\bfseries }&\multicolumn{3}{c}{\bfseries Digital (Call)}\tabularnewline
\cline{2-4} \cline{6-8}
\multicolumn{1}{l}{}&\multicolumn{1}{c}{$\hat{\alpha}$}&\multicolumn{1}{c}{SE}&\multicolumn{1}{c}{$2^{\hat{\alpha}}$}&\multicolumn{1}{c}{}&\multicolumn{1}{c}{$\hat{\alpha}$}&\multicolumn{1}{c}{SE}&\multicolumn{1}{c}{$2^{\hat{\alpha}}$}\tabularnewline
\hline
{\bfseries Incidence ($\hat \alpha^{0\rightarrow 1}$)}&&&&&&&\tabularnewline
~~Current Common Partner&2.867&.006&7.295&&&&\tabularnewline
~~General Common Partner&.726&.007&1.654&&.224&.125&1.168\tabularnewline
~~Number Interaction&1.129&.005&2.187&&1.630&.039&3.095\tabularnewline
~~Friendship Match&.383&.010&&&5.701&.117&\tabularnewline
~~Gender Match&$-$.023&.010&&&.207&.085&\tabularnewline
\hline
{\bfseries Duration ($\hat \alpha^{1\rightarrow 0}$)}&&&&&&&\tabularnewline
~~Current Interaction&$-$.052&.001&.965&&$-$.028&.022&.981\tabularnewline
~~Number Interaction&$-$.154&.005&.899&&$-$.292&.073&.817\tabularnewline
~~Current Common Partner&$-$.318&.006&.802&&&&\tabularnewline
~~General Common Partner&.075&.009&1.053&&.513&.186&1.427\tabularnewline
~~Friendship Match&$-$.556&.010&&&$-$.498&.458&\tabularnewline
~~Gender Match&$-$.017&.010&&&$-$.324&.318&\tabularnewline
\hline
\end{tabular}\end{center}
\end{table}

In the incidence model, shared current common partners are a key determinant of physical interactions. 
The first shared partner between actors $i$ and $j$ increases the corresponding event intensity by a multiplicative factor of $7.295$. 
This coefficient confirms that joining an existing group of people is much more likely than starting a physical interaction with a single actor.
General common partners also affect the event intensities. 
However, their effects are moderate: they increase the incidence intensity of physical interactions by a multiplicative factor of 1.654 and of digital interactions by 1.168 in the same setting.
Besides these triadic closure effects, repeated interactions affect the incidence intensity. 
The first interaction between arbitrary actors $i$ and $j$ increases $\lambda_{i,j}^{0 \rightarrow 1}\left(t\mid \mathscr{H}_t, \bm{\theta}^{0 \rightarrow 1}\right)$ of physical and digital interactions by multiplicative factors of 2.187  and 3.095, respectively. 
By the interpretation shown in \eqref{eq:interpretation_exp}, a shared Facebook friend affects the incidence intensity of digital communication by the multiplicative factor $\exp(5.701) \approx 299$. 
The incidence intensity of physical meetings is increased only by the factor $\exp(.383) \approx 1.46$ in the same setting. 
We estimate divergent effects for gender homophily. 
The incidence intensity of digital interaction between two actors that match on their gender is significantly higher than for pairs of different genders. 
For physical interactions, we observe the opposite effect. 
\hide{
In general, the results of the model for physical interactions based on co-location align with the results obtained in \citet{hoffman_model_2020}, where similar data was studied from a group-level perspective.
}

Both the number of current common partners and prior interactions significantly prolong physical interactions. 
The first current common partner and interaction reduce the duration intensity by multiplicative factors of 0.802 and 0.899, respectively. 
Similarly, shared friendship status approximately halves the duration intensity of physical interactions.
This finding suggests that strong relational ties stabilize over time.
In contrast, gender homophily does not have a statistically significant effect on the duration intensity of physical interactions. 
The durations of digital interactions, measured via call events, are driven by similar factors. 
However, in this case, none of the exogenous effects significantly differ from zero.


\subsection{Model Assessment}
\label{sec:oos}

\begin{figure}[t!]
  \centering
      \includegraphics[width=0.99\textwidth]{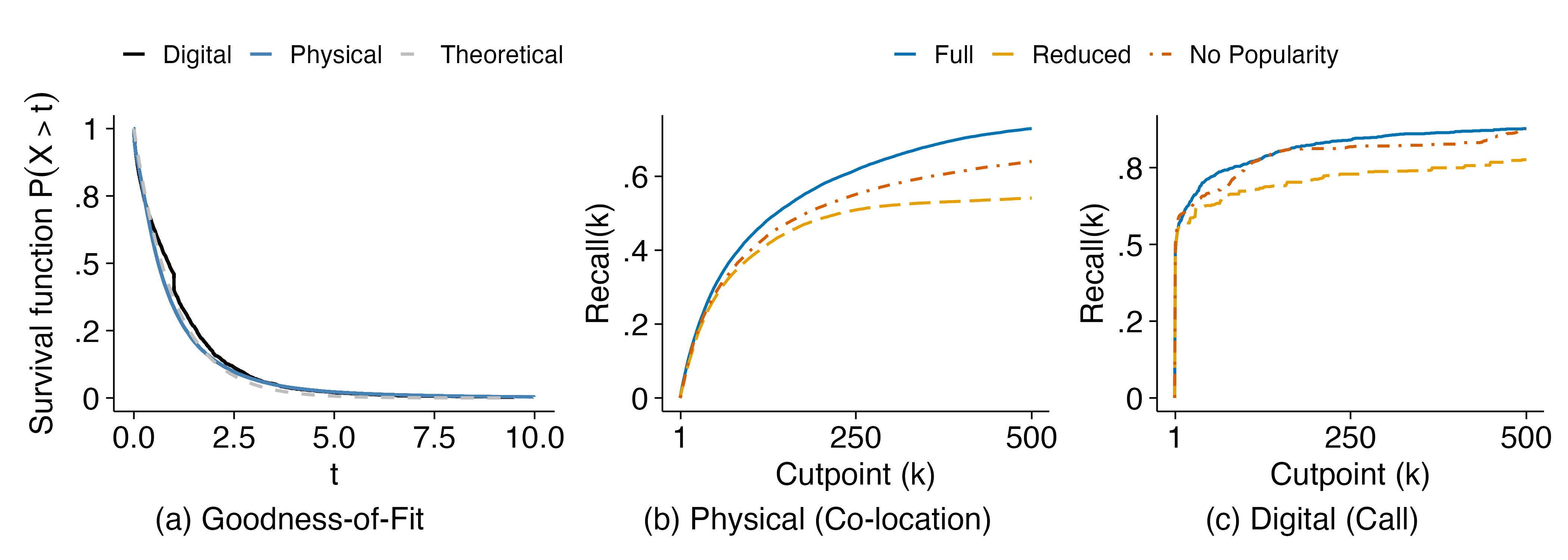}  
    \caption{(a) Comparison of empirical and theoretical survival functions. (b) and (c): Top-$k$ recall curves applied to physical (b) and digital (c) interaction data for three model specifications: Full (blue), Reduced (yellow), and No Popularity Models (orange).}
  \label{fig:gof}
\end{figure}

To ensure that the estimated model adequately fits the observed data, we assess in-sample fit via Cox-Snell residuals and out-of-sample prediction.
The time rescaling theorem \citep{brown2002} establishes for a correctly specified model that integrating all conditional intensities over their respective piecewise constant intervals yields a sequence of independent and identically distributed unit exponential random variables.
These quantities, termed Cox-Snell residuals, are right-censored for the final interval of each dyad. 
Therefore, we evaluate goodness-of-fit by comparing a nonparametric Kaplan-Meier estimator of the survival function, defined as $\mathbb{P}(X > t)$ with $t >0$ for a nonnegative random variable $X$, against the corresponding theoretical value, where $\mathbb{P}(X > t) = \exp(-t)$ holds under $X \sim \text{Exp}(1)$.
As illustrated in Figure \ref{fig:gof} (a), the estimated survival functions for both physical (black) and digital (blue) interactions closely follow the theoretical curve, 
suggesting that the proposed model accurately recovers the temporal structure of the data.

Following \citet{vu_continuous-time_2011}, we assess out-of-sample fit by holding out the final $20\%$ of events in the model for physical and digital interaction. 
With the remaining $80\%$ of the data, we re-estimate three model specifications: 
the full model in Section \ref{sec:specification} (Full), 
a reduced model with only friendship and gender statistics (Reduced), and 
the full model without popularity terms and log-transformed statistics (No Popularity).   
With \eqref{eq:intensity_comb}, the probability of an event between actors $i$ and $j$ to be the first after time $t$ under piecewise constant intensities is:
\be
\label{eq:prob}
\mathbb{P}((i,j) \text{ is next event after } t) &=& \dfrac{\lambda_{i,j}\left(t\mid \mathscr{H}_t, \bm{\theta}\right)}{\sum_{(h,k) \in \actorsubset^\star(t) } \lambda_{h,k}\left(t\mid \mathscr{H}_t, \bm{\theta}\right)}.
\ee
For each event in the last $20\%$ of events, 
we rank the top-$k$ most likely events according to each of the three models and record whether the observed event falls within the respective top-$k$ list. 
After each event, we update the sufficient statistics and iterate this procedure for the next event in the test set. 
The resulting top-$k$ recall measure for $k\in \{1,\ldots\}$ and a model is the fraction of events in the test set within the top-$k$ predictions.
As illustrated in Figure \ref{fig:gof} (b) \& (c), the full specification yields the best predictive power.

\section{Conclusion}
\label{sec:discussion}

\hide{
Traditional frameworks often treat interactions or occurrences as discrete, static events that either exist or do not exist. The innovation of the proposed modeling framework for durational event data lies in its ability to capture the temporal extent of occurrences within relationships, enabling the model to account for the temporal context of interactions. This is crucial, as most real-world networks exhibit temporal dynamics that cannot be fully understood by considering isolated occurrences alone.
}

The proposed framework, accompanied by a scalable estimation algorithm and implementation in the $\mathtt{redeem}$ package, offers a foundation for modeling the dynamics of durational events.
Nonetheless, there are still several avenues for extensions.
For massive populations, it is reasonable to assume that durational events between actors  $i$  and  $j$  are primarily influenced by their respective local neighborhoods, rather than by the entire network. 
This assumption can be represented by local dependence \citep{schweinberger_local_2015}, which can be extended to the context of durational events.
Concerning algorithmic enhancements, using sample-based approximations \citep{rafteryFastInferenceLatent2012} or online algorithms \citep{corneck2025} with our coordinate ascent algorithm will enable scaling to larger populations. 

\section*{Acknowledgments}

The authors are indebted to the constructive comments and suggestions of an anonymous associate editor and three referees, which have led to numerous improvements. 

\section*{Funding}

This publication has emanated from research conducted with the financial support of Taighde \'Eireann – Research Ireland, under Grant number $[23/RC/13506]$ at the Research Ireland Centre - Rinn Artificial Intelligence.

\section*{Replication Materials}
Replication materials are provided in the following GitHub repository:
\begin{center}
    \url{https://github.com/corneliusfritz/replication_dem}
\end{center}
\hide{
Nonetheless, there are still several avenues for future extensions.
Concerning the model specification, the integration of multiple states and higher-order events such as meetings between multiple actors lasting a particular time (see, e.g., \citealp{socio_cognitive_aes_jl_cf}) 
would further broaden the model’s applicability. 
For massive populations, it is reasonable to assume that durational events between actors  $i$  and  $j$  are primarily influenced by their respective local neighborhoods, rather than by the entire network. 
This assumption is grounded in the concept of local dependence of \citet{schweinberger_local_2015}, which can be extended to the context of durational events.

Concerning algorithmic enhancements, the modular structure of the proposed block-coordinate ascent algorithm from Section \ref{sec:computing} allows for the independent application of any approximation or acceleration technique in \eqref{eq:step1}, \eqref{eq:update_MM}, and \eqref{eq:update_gamma} for the three updates separately. 
In particular, it will be worthwhile to adapt sample-based approximations, such as case-control sampling \citep{lerner2020} or Horvitz–Thompson-estimators \citep{rafteryFastInferenceLatent2012}, to scale estimation to even larger networks. 
For Step 3, acceleration techniques tailored to MM algorithms, discussed in \citet{agarwal2024}, can be employed to improve convergence speed.
}

\hide{
Finally, we provide the R package $\mathtt{redeem}$, which implements our method and the suite of summary statistics introduced in Section \ref{sec:summary_statistics}. 
To accommodate problem-specific requirements and different data sources, users can define custom summary statistics and transformations using $\verb!C++!$ code.
Our approach parallels the functionality of the $\mathtt{ergm.userterms}$ package, which enables the inclusion of user-defined statistics for estimating Exponential Random Graph Models \citep{hunterErgmusertermsTemplatePackage2013}. 
As such, $\mathtt{redeem}$ is a flexible and extendable toolbox, allowing researchers to tailor the statistical framework to meet their specific modeling needs.
}

\bibliography{references}

\setcounter{footnote}{0}

\spacingset{1.9} 

\renewcommand{\thefootnote}{\fnsymbol{footnote}} 
\newpage
\setcounter{page}{1}
\appendix
\begin{center}
	{\LARGE\textbf{Supplementary material\\  Scalable Durational Event Models: Application to Physical and Digital Interactions} } \\ 
	\hspace{.2cm}\\
	\vspace{0.3cm}
    \if1\anon
{Cornelius Fritz$^1\footnote{\textbf{Corresponding author:} \href{mailto:fritzc@tcd.ie}{fritzc@tcd.ie}.
}$, Riccardo Rastelli$^{2,3}\footnote{
Riccardo Rastelli, Michael Fop, and Alberto Caimo contributed equally to the work. Alberto Caimo conceived the original idea of the model.
}$, Michael Fop$^2\footnotemark[2]$,  Alberto Caimo$^2\footnotemark[2]$\\[.2cm] $^1$School of Computer Science and Statistics, Trinity College Dublin  \vspace{.1cm}\\ $^2$School of Mathematics and Statistics, University College Dublin \\
$^3$Rinn Artificial Intelligence, University College Dublin, Ireland
	\hspace{.2cm}
}\fi

\end{center}

\numberwithin{equation}{section}
\renewcommand{\cftsecfont}{\mdseries}
\renewcommand{\cftsecpagefont}{\mdseries}
\setlength{\cftbeforesecskip}{0 pt} 
\renewcommand{\cftsecleader}{\cftdotfill{\cftdotsep}} 
\startcontents
\printcontents{ }{1}{}
\newpage

\renewcommand{\thefootnote}{\arabic{footnote}} 
\setcounter{equation}{0}

\section{Further Information on the Simulation Study}
\label{sec:simulation_add}

\hide{
\subsection{Simulation Setup}
\label{sec:setup}

The continuous covariate  $\bm x_1 = (x_{1,1}, \ldots, x_{N,1})$  is drawn from a standard normal distribution, i.e.,  $X_{i,1} \overset{\text{i.i.d}}{\sim} \mathcal{N}(0,1)$  independently for  $i = 1, \dots, N$. 
The discrete covariate  $\bm x_2 = (x_{1,2}, \ldots, x_{N,2})$  is sampled independently from a categorical distribution with three equally probable outcomes.
Three summary statistics are chosen for the incidence and duration: $\bm{s}_{i,j}^{0 \rightarrow 1}(\mathscr{H}_t) = (s_{i,j,CCP}(\mathscr{H}_t),\ s_{i,j,\bm{z}_1}(\mathscr{H}_t),\ s_{i,j,\bm{z}_2}(\mathscr{H}_t) )^\top$ and  $\bm{s}_{i,j}^{1 \rightarrow 0}(\mathscr{H}_t) = (s_{i,j,NI}(\mathscr{H}_t),\ s_{i,j,\bm{z}_1}(\mathscr{H}_t),\ s_{i,j,\bm{z}_2}(\mathscr{H}_t))^\top$,
where $z_{i,j,1} = |x_{i,1} - x_{j,1}|$ and $z_{i,j,2} = \mathbb{I}(x_{i,2} = x_{j,2})$.
The true parameters are as follows: 
\begin{itemize}
    \item $\bm{\alpha}^{\,0\rightarrow 1} = (-1/2,1,1/2)$ and $\bm{\alpha}^{1\rightarrow 0} = (1/2,1/2,1/2)$;
    \item 
    The popularity parameters decrease on average with increasing $N$, implying sparsity of observed events as the number of actors $N$ grows:
\beno 
{\beta}^{0 \rightarrow 1}_i &\overset{\text{i.i.d}}{\sim}& \mathcal{N}(-6-1/10\,\log(N),1) \\ \beta^{1 \rightarrow 0}_i &\overset{\text{i.i.d}}{\sim}& \mathcal{N}(8/5-1/10\,\log(N),1).
\ee 
\item We set the time interval as $\mathscr{T} = [0,\mbox{10,000}]$ 
with nine equally spaced change points $c_1, ..., c_9$. 
The true values of the baseline function, $\bm{\gamma}^{0\rightarrow 1} = \bm{\gamma}^{1\rightarrow 0}$,  are linearly decreasing from zero to $-1/10$. 
\end{itemize}
}

\subsection{Simulation of Durational Events} 
\label{sec:simulation}

As an alternative to the canonical representation of an undirected durational event between the pair of actors $(i,j) \,\in\, \dyadset$, beginning at time $t_b \in \timeset$ and ending at time $t_e \in \timeset$, with $t_b<t_e$, by a four-dimensional tuple, $d = (i,\,j,\,t_b,\,t_e)$, we can reformulate this data as follows: 
a durational event is characterized by two tuples in the form $(i,j,t,r)$, where $t$ is a time stamp, and $r$ is a binary indicator that determines if the tuple refers to the beginning ($r=1$) or end ($r=0$) of an interaction. 
Then we can equivalently represent durational event $d= (i,j,t_b,t_e)$ by ${d}_{\text{start}} = (i,j,t_b,1)$ and $d_{\text{end}} = (i,j,t_e,0)$.
This is the format used for the provided R package $\mathtt{redeem}$ and in the following description of the pseudo-code to sample durational events from (2). 

We present Algorithm \ref{al:simulation} for sampling durational events within an arbitrary time frame $(0, T]$ under the assumption that $f(t,\bm \gamma^{0\rightarrow 1}) = f(0,\bm \gamma^{0\rightarrow 1})$ and $f(t,\bm \gamma^{1\rightarrow 0}) = f(0,\bm \gamma^{1\rightarrow 0})$ remain constant in that interval. 
When these baseline step functions vary over time as step functions, we apply Algorithm \ref{al:simulation} separately to each interval and concatenate the results.
To accommodate for a flexible specification of $\actorsubset^{0 \rightarrow 1}(t)$ and $\actorsubset^{1 \rightarrow 0}(t)$, we include for each pair $(i,j) \in \dyadset$ an indicator $p_{i,j}(t)\in \{0,1\}$ of whether any type of event, start or end, is possible between the actors at time $t\in \mathscr{T}$.  
Our algorithm, therefore, assumes the specification of the following terms: 
\begin{enumerate}[leftmargin=1cm]
    \item Number of actors $N$ in the durational event network.
    \item Sets of summary statistics for the incidence and duration model, $\bm s_{i,j}^{0 \rightarrow 1}(\mathscr{H}_t)$ and $\bm s_{i,j}^{1 \rightarrow 0}(\mathscr{H}_t)$.
    \item Parameter vectors $\bm \alpha^{0 \rightarrow 1}$ and $\bm \alpha^{1 \rightarrow 0}$ for the summary statistics. 
    \item Parameter vectors $\bm \beta^{0 \rightarrow 1}$ and $\bm \beta^{1 \rightarrow 0}$ for the popularity effects.
    \item Parameters $f(0, \bm{\gamma}^{0\rightarrow 1})$ and $f(0, \bm{\gamma}^{1\rightarrow 0})$. 
\end{enumerate}
Together, these parameters are defining $\bm{\theta}^{0\rightarrow 1}$  and $\bm{\theta}^{1\rightarrow 0}$.
The binary indicator tracking whether dyad $(i,j) \in \dyadset$ is in an ongoing interaction at time $t \in \timeset$ is given by $a_{i,j}(t)$.
We follow the convention to denote observed values of a random variable $X$ by $x$ and the resulting durational event matrix $E\in \mathbb{R}^{L \times 4}$ includes in each row one sampled event that can either be the beginning or end of a durational event.

\begin{algorithm}
 \renewcommand{\baselinestretch}{1.3}\normalsize
    \SetAlgoLined
	\KwResult{Durational Event Matrix $E \in \mathbb{R}^{L \times 4}, \quad E_{l,1}, E_{l,2} \in \actorset, \quad E_{l,3} \in \mathscr{T},$ $ \quad E_{l,4} \in \{0,1\}$ for $l \in \{1, ..., L\}$, where $L$ is the maximal number of events. }
	Set $\tilde{t} = 0$ and $\tilde{n} = 1$. \\
	\While{$\tilde{t} < T$ and $\tilde{n}\leq L$}{
		 \textbf{1. Step: Compute Intensities} \\  
         {Set}          
         \beno 
         \lambda_{i,j}^{0 \rightarrow 1}\left(\tilde{t}\mid \mathscr{H}_{\tilde{t}}, \bm{\theta}^{0 \rightarrow 1}\right) &=&\exp\left(\bm \alpha^{0 \rightarrow 1}\, \bm s_{i,j}^{0 \rightarrow 1}(\mathscr{H}_{\tilde{t}}) + \beta^{0 \rightarrow 1}_i +\beta^{0 \rightarrow 1}_j + f(0, \bm{\gamma}^{0\rightarrow 1})\right)
         \ee
         and 
         \beno 
         \lambda_{i,j}^{1 \rightarrow 0}\left(\tilde{t}\mid \mathscr{H}_{\tilde{t}}, \bm{\theta}^{1 \rightarrow 0}\right)
         &=&\exp\left(\bm \alpha^{1 \rightarrow 0}\, \bm s_{i,j}^{1 \rightarrow 0}(\mathscr{H}_{\tilde{t}}) + \beta^{1 \rightarrow 0}_i +\beta^{1 \rightarrow 0}_j + f(0,\bm{\gamma}^{1\rightarrow 0})\right)
         \ee
         for all $(i,j)\in \dyadset$.
         
      \textbf{2. Step: Select Intensities} \\
      {Set} $\bm{\lambda} \left(\tilde{t}\mid \mathscr{H}_{\tilde{t}}, \bm{\theta}^{0 \rightarrow 1}, \bm{\theta}^{1 \rightarrow 0}\right) = \left(\lambda_{i,j}\left(\tilde{t}\mid \mathscr{H}_{\tilde{t}}, \bm{\theta}^{0 \rightarrow 1}, \bm{\theta}^{1 \rightarrow 0}\right)\right)$
         \beno 
         \lambda_{i,j}\left(\tilde{t}\mid \mathscr{H}_{\tilde{t}}, \bm{\theta}^{0 \rightarrow 1}, \bm{\theta}^{1 \rightarrow 0}\right) &=& \big((1-a_{i,j}(\tilde{t}))\, \lambda_{i,j}^{0 \rightarrow 1}\left(\tilde{t}\mid \mathscr{H}_{\tilde{t}}, \bm{\theta}^{0 \rightarrow 1}\right)\s \\ &+& a_{i,j}(\tilde{t})\, \lambda_{i,j}^{1 \rightarrow 0}\left(\tilde{t}\mid \mathscr{H}_{\tilde{t}}, \bm{\theta}^{1 \rightarrow 0}\right) \big)\, p_{i,j}(\tilde{t})
         \ee
        for all $(i,j)\in \dyadset$.

      \textbf{3. Step: Sample Time Increment} \\  
      {Sample} time between successive events from:
      \beno 
      T^\star &\sim& \text{Exp}(\zeronormvec{\bm{\lambda} \left(\tilde{t}\mid \mathscr{H}_{\tilde{t}}, \bm{\theta}^{0 \rightarrow 1}, \bm{\theta}^{1 \rightarrow 0}\right)}).
      \ee

      \textbf{4. Step: Sample Event} \\  
      {Sample} which pair will experience the event from: 
      \beno 
      (I^\star,J^\star) &\sim& \text{Multinomial}\left(\dfrac{\bm{\lambda} \left(\tilde{t}\mid \mathscr{H}_{\tilde{t}}, \bm{\theta}^{0 \rightarrow 1}, \bm{\theta}^{1 \rightarrow 0}\right)}{\zeronormvec{\bm{\lambda} \left(\tilde{t}\mid \mathscr{H}_{\tilde{t}}, \bm{\theta}^{0 \rightarrow 1}, \bm{\theta}^{1 \rightarrow 0}\right)}}, n = 1\right).
      \ee 
      \\  
      \textbf{5. Step: Save Sampled Event from} \\
      {Set}  $ E_{\tilde{n},1} = i^\star, E_{\tilde{n},2} = j^\star, E_{\tilde{n},3} = t^\star+\tilde{t},$ and $ E_{\tilde{n},4} = 1-a_{i^\star,j^\star}(\tilde{t})$. 
      \\  \textbf{6. Step: Update History and Counters} \\
      {Update} $\bm s_{i,j}^{0 \rightarrow 1}(\mathscr{H}_{t^\star+\tilde{t}})$ and $\bm s_{i,j}^{1 \rightarrow 0}(\mathscr{H}_{t^\star+\tilde{t}})$ for all $(i,j)\in \dyadset$. \\
      {Set} $a_{i^\star,j^\star}(\tilde{t}+ t^\star) = 1 - a_{i^\star,j^\star}(\tilde{t}), \tilde{t} = \tilde{t}+t^\star,$ and $\tilde{n} = \tilde{n} + 1$. 
	}
    \textbf{Return } $E[1:(\tilde{n}-1),1:4]$
    \caption{Pseudo-Code to sample durational events in $(0,T]$.}
	\label{al:simulation}
\end{algorithm}

\subsection{Definition of Performance Measures}
\label{sec:fruther_simulation}
The Average Estimate (AVE) of estimate $\hat {\bm{\theta}}$ is the empirical mean of the estimates over the $S$ replications of the simulation study: 
\beno 
    \text{AVE}(\hat{\bm{\theta}}) &=& \dfrac{1}{S} \dsum_{s = 1}^S \hat{\bm{\theta}}_{(s)},
\ee
where $\hat{\bm{\theta}}_{(s)} = (\hat{\bm{\alpha}}_{(s)}, \hat{\bm{\beta}}_{(s)}, \hat{\bm{\gamma}}_{(s)})$ is the estimate of $\bm{\theta}$ in the $s$th simulation run. 
Given that all datasets are generated under the same parameter values, this statistic provides a meaningful assessment of estimation bias via
\beno 
    \text{Bias}(\hat{\bm{\theta}}) &=& \text{AVE}(\hat{\bm{\theta}}) - \bm{\theta}^*, 
\ee
where $\bm{\theta}^*$ is the data-generating parameter. 
To gauge both bias and variance, we report the Root Mean-Squared Error (RMSE) of $\hat{\bm{\theta}}$: 
\beno 
   \text{RMSE}(\hat{\bm{\theta}}) &=&  \sqrt{\dfrac{1}{S} \dsum_{s = 1}^S\left(\hat{\bm{\theta}}_{(s)} - \bm{\theta}^* \right)^\top \left(\hat{\bm{\theta}}_{(s)}- \bm{\theta}^*\right)}.
\ee

To assess uncertainty quantification when estimating $\bm{\alpha}$, we examine coverage probabilities.
The coverage probability is the percentage of simulations in which the true parameter falls within the confidence interval based on the normal approximation: 
\beno
\label{eq:normal_approx}
\bm{Z}_{(s)} \,=\, \bm{\Lambda}(\hat{\bm{\theta}}_{(s)})^{1/2}(\hat{\bm{\alpha}}_{(s)} - \bm{\alpha}^*) &\sim& \bm{\mathcal{N}}_P(\bm{0}_P, \bm{I}_P), 
\ee
where the block of the inverse Hessian corresponding to $\bm{\alpha}$, $\bm{\Lambda}(\hat{\bm{\theta}})$, is defined in (14) of the main article, $\bm{0}_P = (0,..., 0)^\top \in \mathbb{R}^P$ and $\bm{I}_P \in \{0,1\}^{P\times P}$ denotes the identity matrix.
To validate this normal approximation, we compare the observed quantiles of $\bm{Z}_{(s)}$ to those we expect from standard normal random variables.

\subsection{Further Results}

\subsubsection{Simulation Study 1} 
\label{sec:simulation_1_further}
In addition to assessing the coverage probability and accuracy of the parameter estimates demonstrated in Simulation Study 1 in the main manuscript, we assessed the effectiveness of model selection via the AIC and looked at Quantile-Quantile plots of the standardized estimates. 

\paragraph*{Model Selection.} 
The Akaike Information Criterion (AIC, \citealpsupp{akaikeInformationTheoryExtension1973}) for the DEM with converged estimates $\hat{\bm{\theta}} = ({\hat {\bm{\theta}}^{0\rightarrow 1}}, {\hat{\bm{\theta}}^{1\rightarrow 0}})$ is given by the sum of the AIC of the incidence and duration model: 
\be
\label{eq:aic}
AIC^{\,0\rightarrow 1} &\coloneqq& 2\, P^{0\rightarrow 1} -  2\, \ell({\hat{\bm{\theta}}^{0\rightarrow 1}}) \text{  and  } AIC^{\,1\rightarrow 0} &\coloneqq& 2\,  P^{1\rightarrow 0}  -  2\, \ell({\hat{\bm{\theta}}^{1\rightarrow 0}}), 
\ee
where $P^{0\rightarrow 1}$ and $P^{1\rightarrow 0}$ denote the total number of parameters in the incidence and duration model, respectively. 
We apply the following greedy model selection procedure: first, we fit a simple model in which the incidence model includes only the current common partner, and the duration model has no covariates. 
Then we compute the AIC using \eqref{eq:aic} for this initial model. 
We proceed by adding covariates to the incidence model in a step-wise fashion, one at a time. 
For each fitted model, we again calculate the AIC and retain the new model as optimal if it yields a lower AIC. 
Once we converge to an optimal incidence model structure, we apply an analogous procedure to the duration model. 
The model that achieves the minimum value of the criterion overall is the selected model. 
The inferred optimal model coincides with the data-generating model across all replications in this experiment. 

\begin{figure}[t!]
  \centering
  \includegraphics[width=0.85\textwidth]{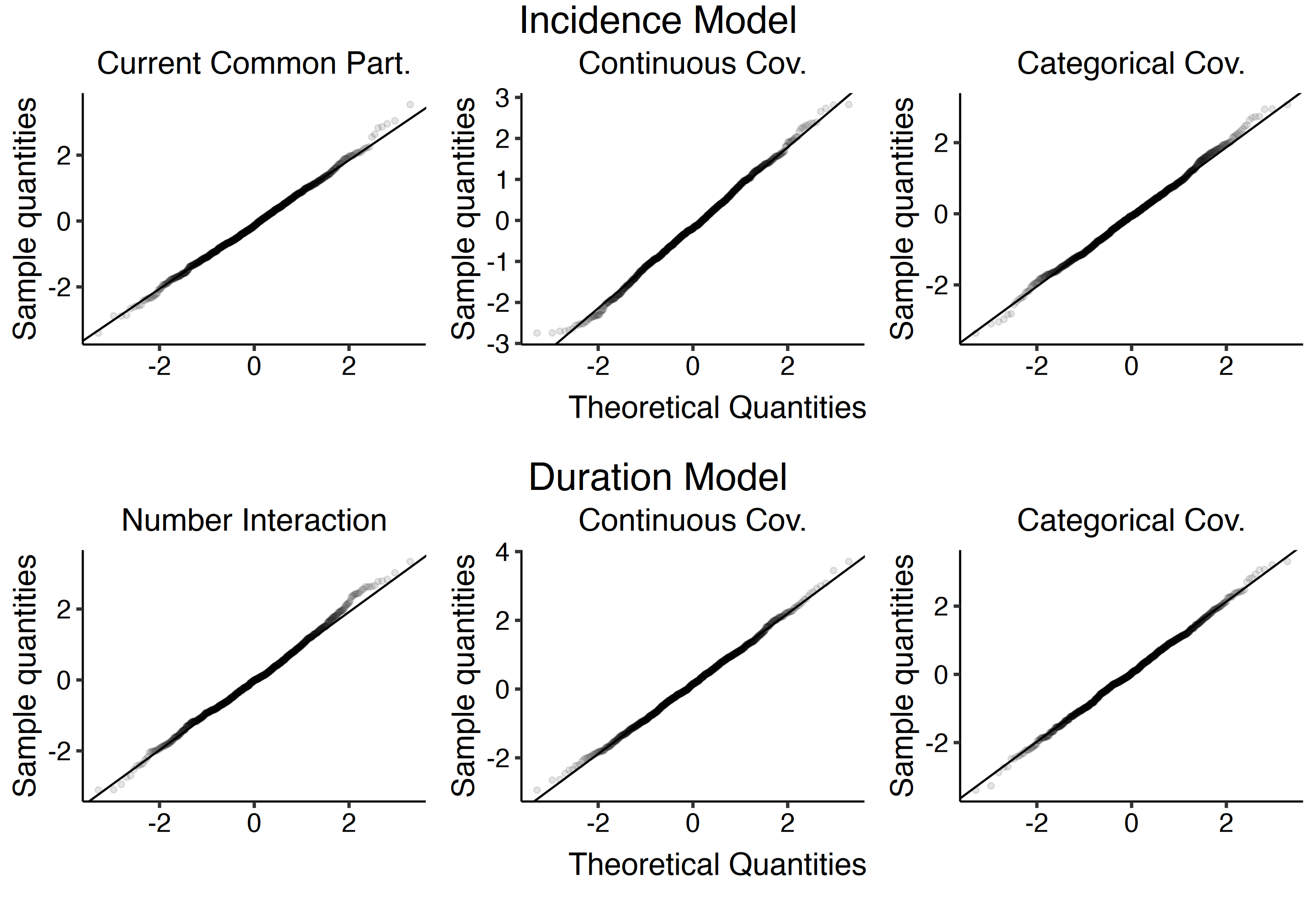}  
    \caption{Simulation Study 1: Quantile-quantile plots for the incidence (top row) and duration model (bottom row) comparing the theoretical quantiles of a standard normal distribution with the sample quantiles of $\bm{z}_{(s)} = \bm{\Lambda}(\hat{\bm{\theta}}_{(s)})^{1/2}(\hat{\bm{\alpha}}_{(s)} - \bm{\alpha}^\star)$ for $s = 1, ..., \mbox{1,000}$. }
  \label{fig:qq_plot}
\end{figure}

\paragraph*{Quantile-Quantile Plots of Standardized Estimates.} 
The quantile-quantile plots in Figure \ref{fig:qq_plot}  illustrate a close alignment of sample quantiles with theoretical normal quantiles, indicating that the normal approximation remains accurate in finite-sample settings.

\subsubsection{Simulation Study 4: Computational Complexity with $M$} 
\label{sec:simulation_4}

Next, we investigate the performance of our estimator as $M$ increases. 
Similar to Simulation Study 1, we fix $N = 500$. However, we set the baseline intensity to be a constant and sample an increasing number of events from this model. 
For each estimation run, we measure the execution time (in seconds) and peak memory allocation (in megabytes) for estimating the model. 

\begin{figure}[t!]
  \centering
  \includegraphics[width=0.95\textwidth]{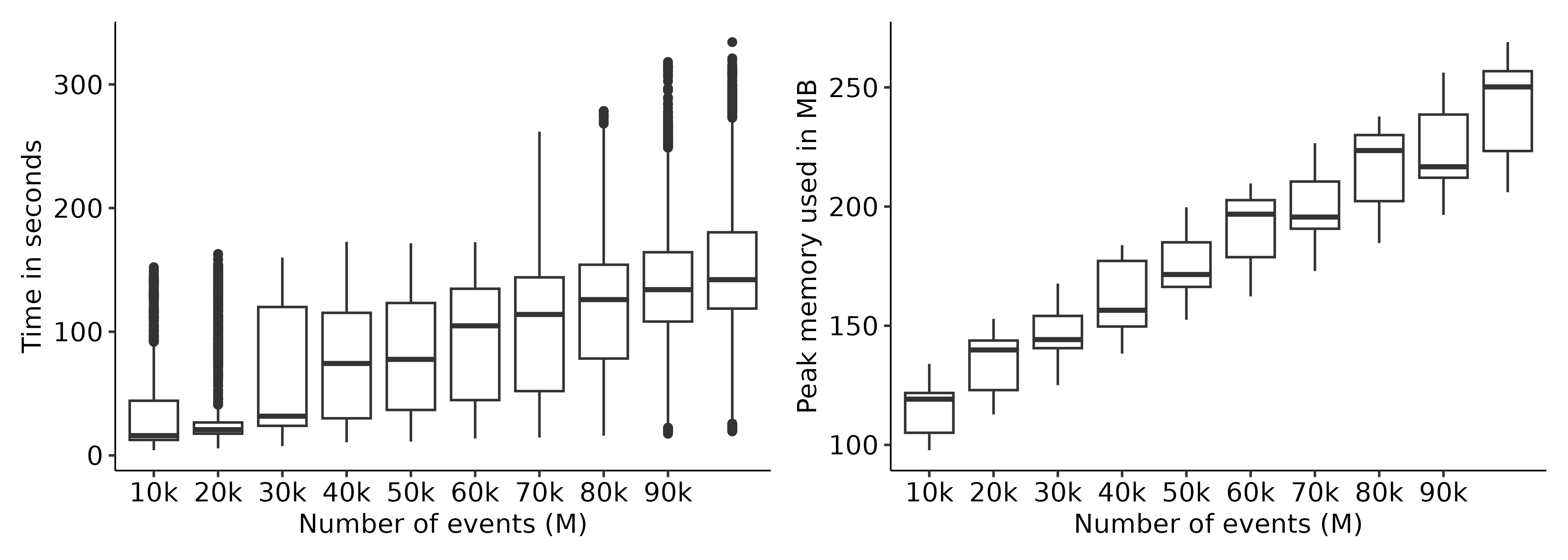} 
    \caption{Simulation Study 4: Comparison of computational time (left) and memory needed for estimation  (right) for different numbers of events $(M)$ of our proposed three-step estimator. }
  \label{fig:sim_4}
\end{figure}
Figure \ref{fig:sim_4} shows a linear growth both in execution time and peak memory used in terms of $M$.
These results align with the provided complexity analysis from Section 3 and show how settings with larger $M$ are not an issue for our algorithm. 

\subsubsection{Simulation Study 5: Choice of $Q$} 
\label{sec:simulation_5}

\begin{figure}[t!]
  \centering
  \includegraphics[width=0.9\textwidth]{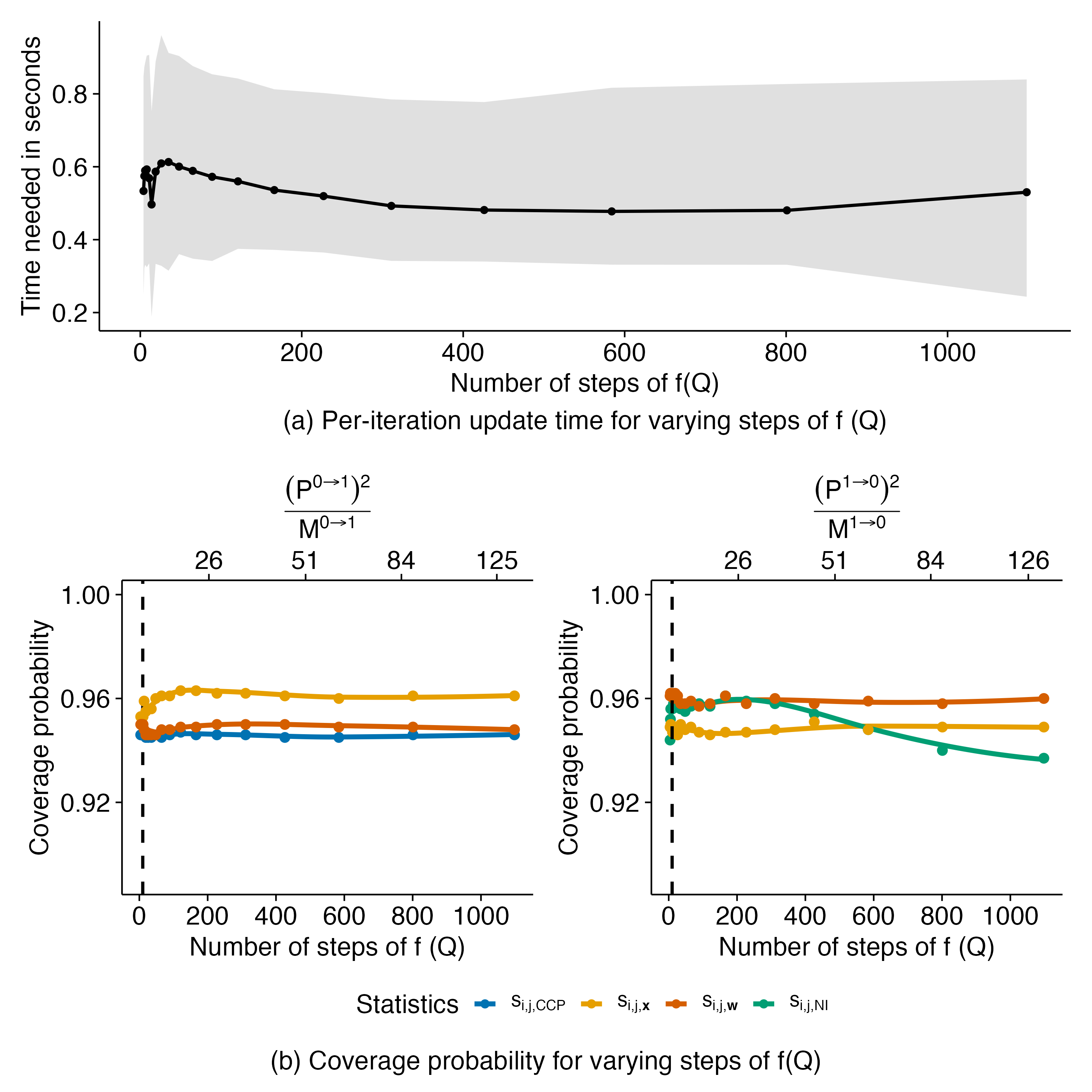} 
    \caption{Simulation Study 5: Coverage probabilities of parameter of interest, $\bm{\alpha}$, under misspecified $Q$. 
    The variables $p^{0\rightarrow 1}$ and $p^{1\rightarrow 0}$ denote the total number of parameters of the incidence and duration model, respectively, with corresponding number of starting and ending events given by $M^{0\rightarrow 1}$ and $M^{1\rightarrow 0}$. 
    }
  \label{fig:sim_5}
\end{figure}

To estimate the DEM, one must choose the number of steps in the baseline intensity, $f(t,\bm{\gamma})$. 
Most times, we regard the parameters $\bm{\gamma}$ characterizing $f$ as nuisance parameters when assessing the uncertainty of the parameters of interest, $\bm{\alpha}$. 
In Simulation Study 5, we assess  two things: 
\begin{enumerate}
    \item The sensitivity of the coverage probabilities of $\bm{\alpha}$ to misspecifying $Q$. 
    \item The computational complexity with varying $Q$. 
\end{enumerate}
Using the same simulation setup as in Simulation Study 1 from Section 4, we re-estimate the model assuming baseline intensities and vary $Q$ between 4 and 1098 in exponential steps, i.e., 4, 5, 6, ..., 584, 801, 1098. 
As with the other simulation studies, we repeat this simulation \mbox{1,000} times. 

In Figure \ref{fig:sim_5} (a), we show that the linear growth of the computational complexity with respect to $Q$ derived in Section 3 broadly aligns with the empirical observed behavior.
Figure \ref{fig:sim_5} (b) illustrates that the coverage probabilities corresponding to all sufficient statistics are relatively constant around $95\%$ with respect to the exact value of $Q$.

On the second upper x-axis of  Figure \ref{fig:sim_5} (b), we plot the squared number of estimated parameters over the number of events beginning (left) and ending (right) an interaction.
The variables $p^{0\rightarrow 1}$ and $p^{1\rightarrow 0}$ thus denote the total number of parameters in the incidence and duration model, respectively. 
First, this provides evidence that the precise choice of $Q$ does not affect the estimator's accuracy or the uncertainty quantification. 
Second, the normal approximation still holds as the ratio of the squared number of parameters to the number of events diverges. 
These empirical results suggest that profiling out $\bm{\gamma}$ and making use of the theory of profile likelihoods still yields valid normal approximations as long as $P$ and $N$ do not grow faster than the square root of the respective number of events \citepsupp{murphy2000}.


\section{Further Information on the Application}

\subsection{Initial Censoring}
\label{sec:initial}
For our applications, we assume that the observations start with the first event. Chronologically ordering the observed durational events by $d_1, \ldots, d_M$, we thus set $d_1 = (i_1, j_1, 0, e_1)$. 
We have no information on when exactly the observational process starts and, therefore, condition on the start of the first event without modeling it. 
 While this approach overlooks the initial censoring of observations, the effect is negligible, since there are $\mbox{155,316}$ and $\mbox{4,152}$ proximity and call events, respectively, which outweigh the contribution of the first event. 
\subsection{Descriptive Plots}
\label{sec:plots}

\begin{figure}[t!]
  \centering
  \begin{subfigure}[c]{0.4\textwidth}   
    \centering
    \includegraphics[width=\textwidth]{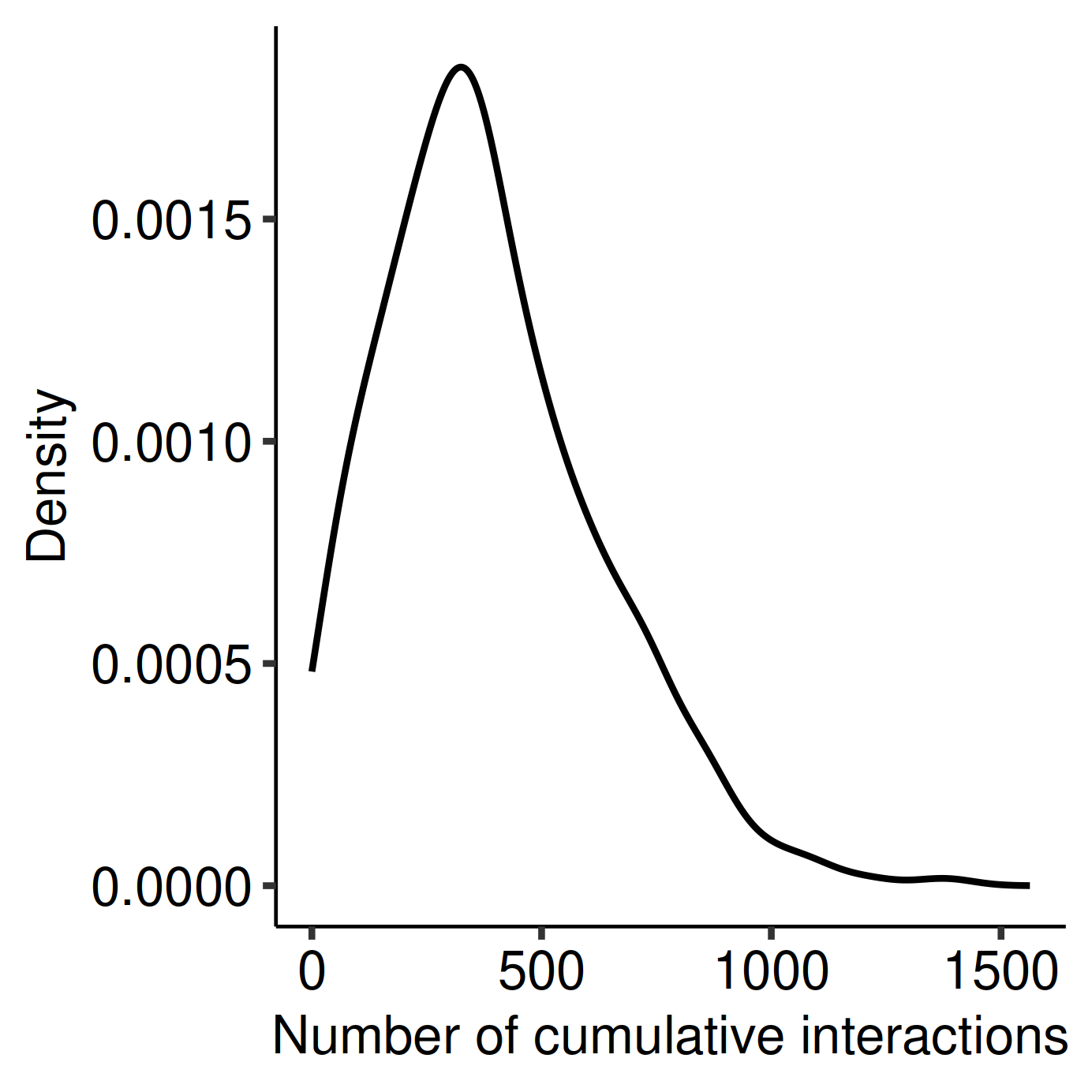}  
    \caption{\hangindent=0em Physical (Co-location)}
  \end{subfigure}
  \begin{subfigure}[c]{0.4\textwidth}  
    \centering
    \includegraphics[width=\textwidth]{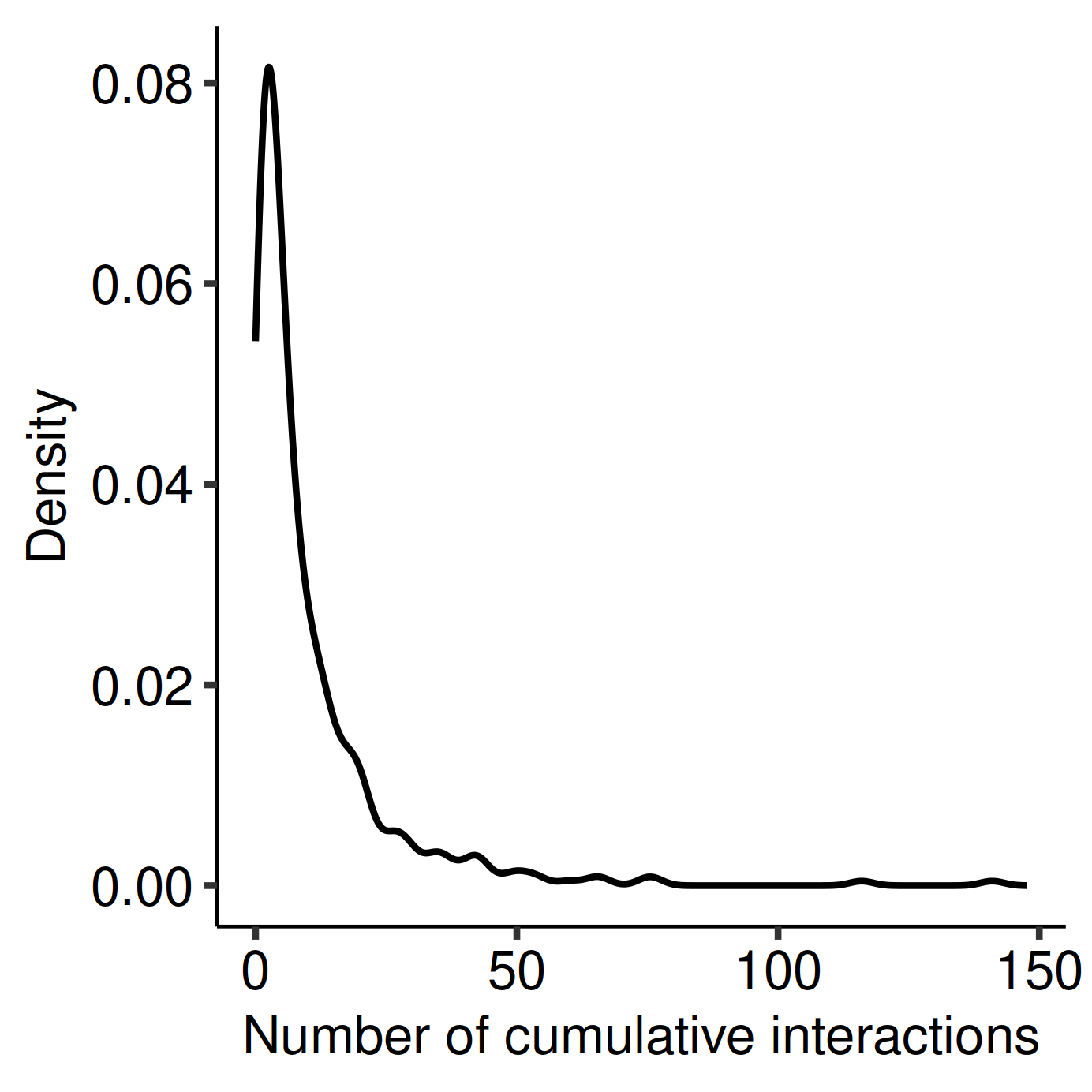}  
    \caption{\hangindent=0em Digital (Call)}
  \end{subfigure}
     \caption{Application: Kernel-density estimator of cumulative durational events per actors for physical (a) and digital interactions (b).}
    \label{fig:degree_stats}
\end{figure}

Figure \ref{fig:degree_stats} presents density estimators of the number of interactions individual actors are involved in for each event network. 
These distributions indicate substantial heterogeneity among actors in both settings and give rise to incorporating popularity coefficients.

\subsection{Popularity and Baseline Estimates}
\label{sec:further_res}

\begin{figure}[t!]
  \centering
  \begin{subfigure}[c]{0.48\textwidth}   
    \centering
    \includegraphics[width=0.9\textwidth]{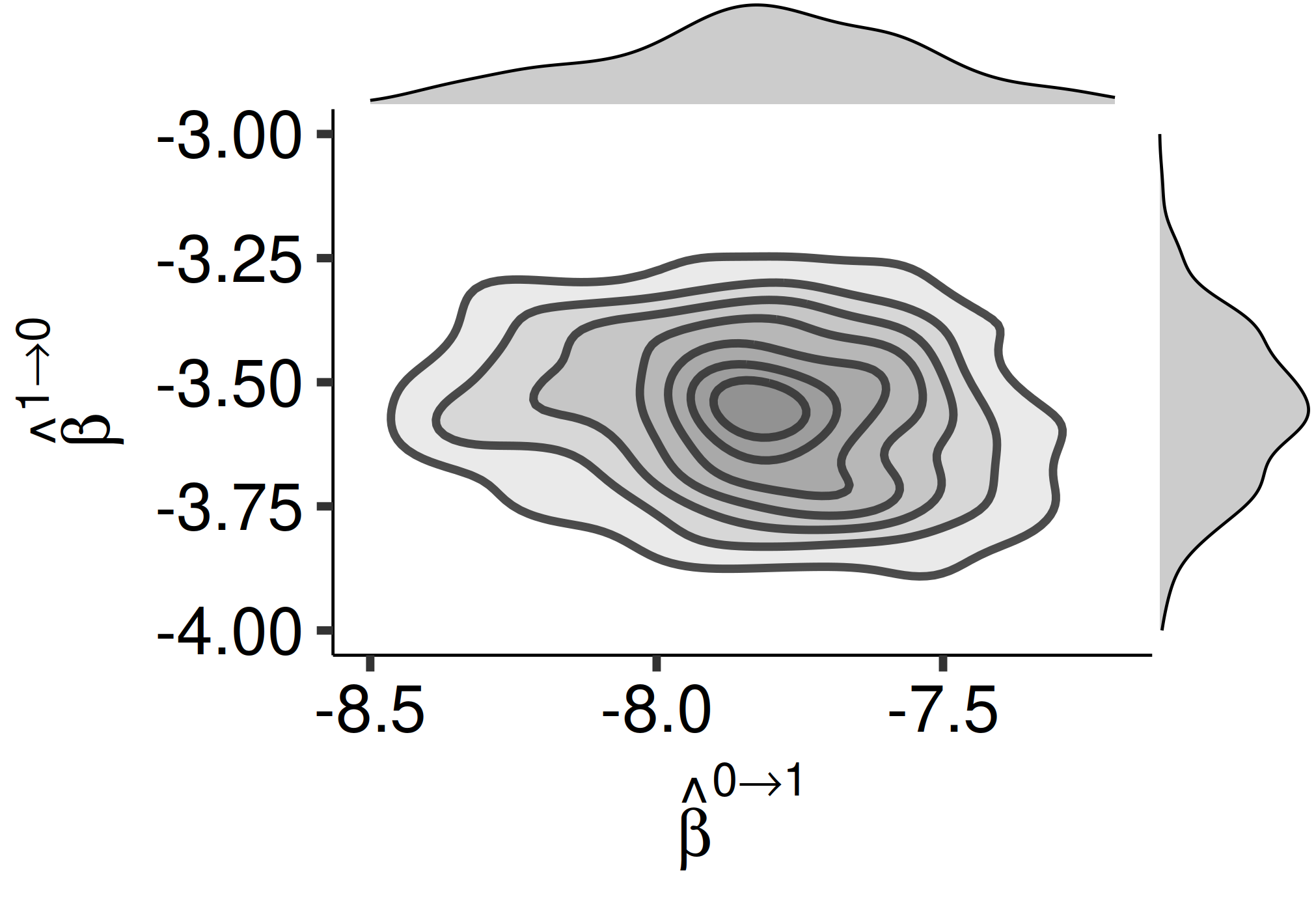}  
    \caption{\hangindent=0em Physical (Co-location)}
    \label{fig:subfig1}
  \end{subfigure}
  \begin{subfigure}[c]{0.48\textwidth}  
    \centering
    \includegraphics[width=0.9\textwidth]{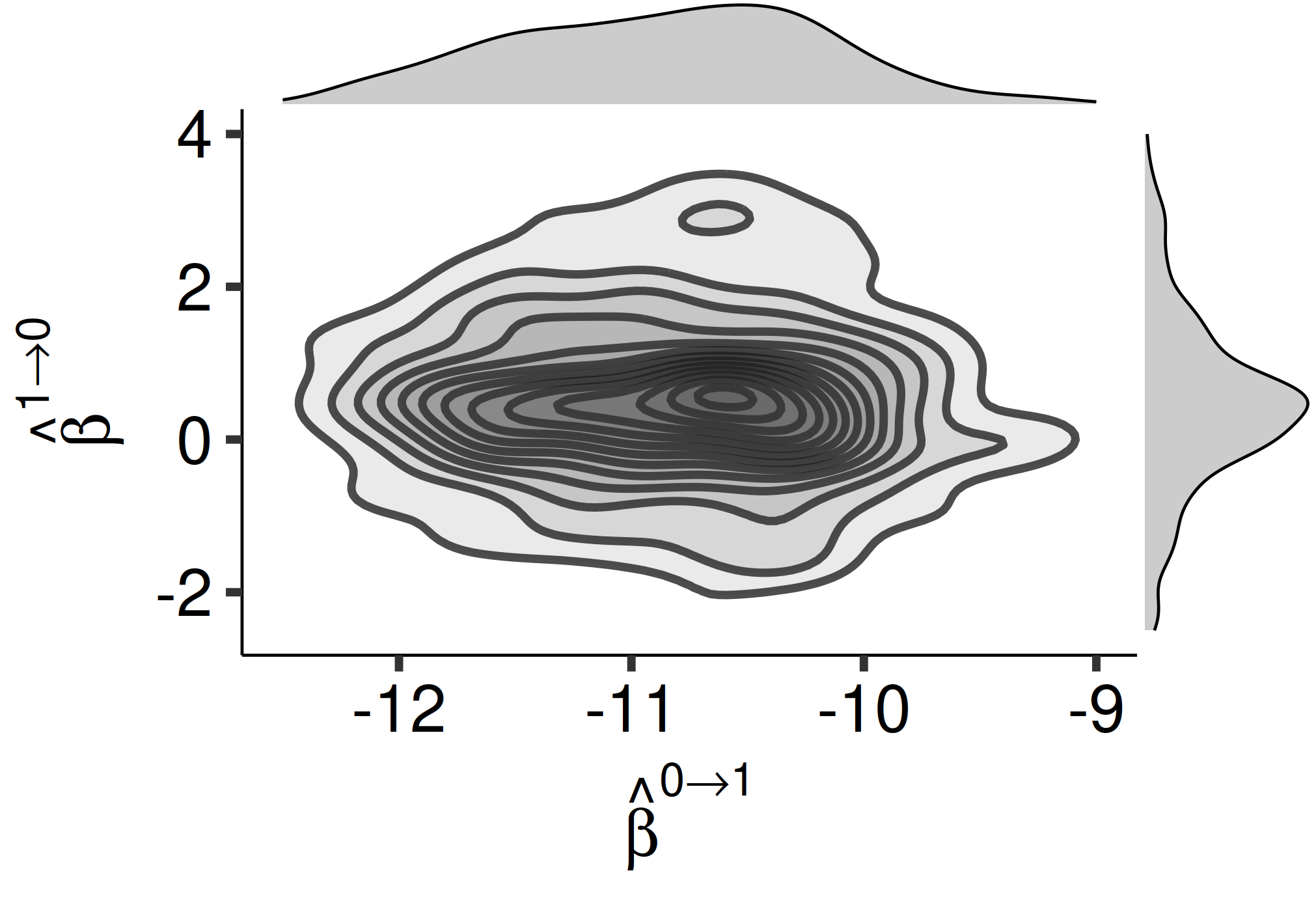}  
    \caption{\hangindent=0em Digital (Call)}
    \label{fig:subfig2}
  \end{subfigure}
    \caption{Two-dimensional kernel density estimates of the popularity parameters for physical (a) and digital interactions (b).}
  \label{fig:est_pop}
\end{figure}

\paragraph*{Popularity Estimates.}
The number of potential actor pairs  to begin interacting is typically orders of magnitude larger than the pairs that may end them, while the number of started and ended events is roughly the same. 
This imbalance is reflected in the overall level of estimates shown in Figure \ref{fig:est_pop}, where the average of $\hat{\bm{\beta}}^{0\rightarrow 1}$ is lower than the average of  $\hat{\bm{\beta}}^{1\rightarrow 0}$ for both interaction modes.
Consistent with Figure \ref{fig:degree_stats}, the popularity parameters in the incidence model are generally higher for co-location than for call interactions. 
The contour lines in Figure \ref{fig:est_pop} enable a comparison of individual actors' popularity estimates between the incidence and duration models.
There is no clear correlation between the two estimates for physical interactions.
However, a pattern emerges for digital interactions where low to average popularity coefficients in the incidence model often correspond to average popularity effects in the duration model.

\paragraph*{Baseline Temporal Estimates.}
If no interaction is observed during a particular hour, we assume that the baseline did not change from the previous hour. Therefore, the number of parameters in $\bm{\gamma}$ fluctuates slightly between each estimated model.

\begin{figure}[t]
  \centering
  \begin{subfigure}[c]{0.48\textwidth}   
    \centering
    \includegraphics[width=\textwidth]{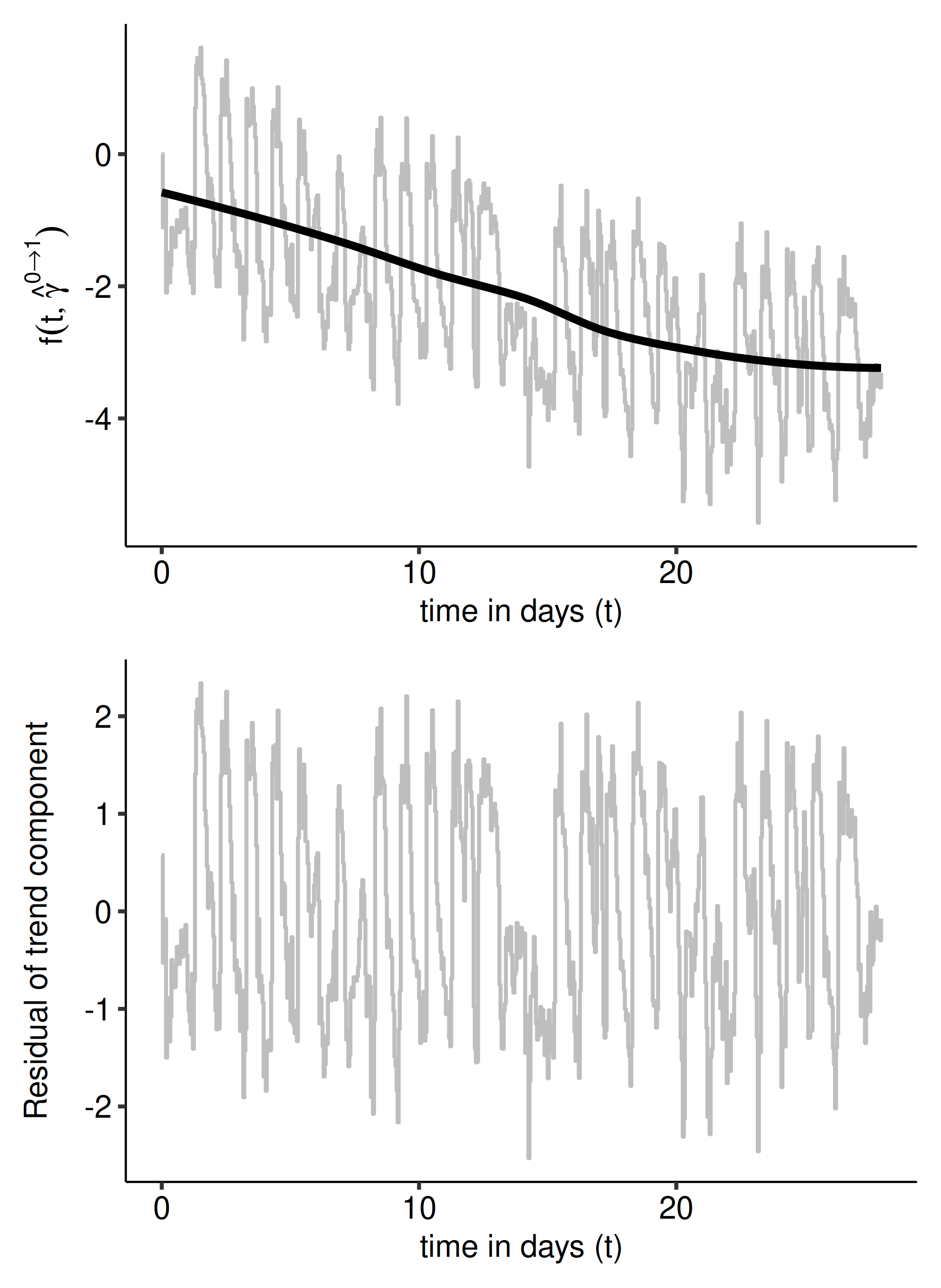} 
    \caption{\hangindent=0em Incidence}
    \label{fig:subfig1}
  \end{subfigure}
  \begin{subfigure}[c]{0.48\textwidth}  
    \centering
    \includegraphics[width=\textwidth]{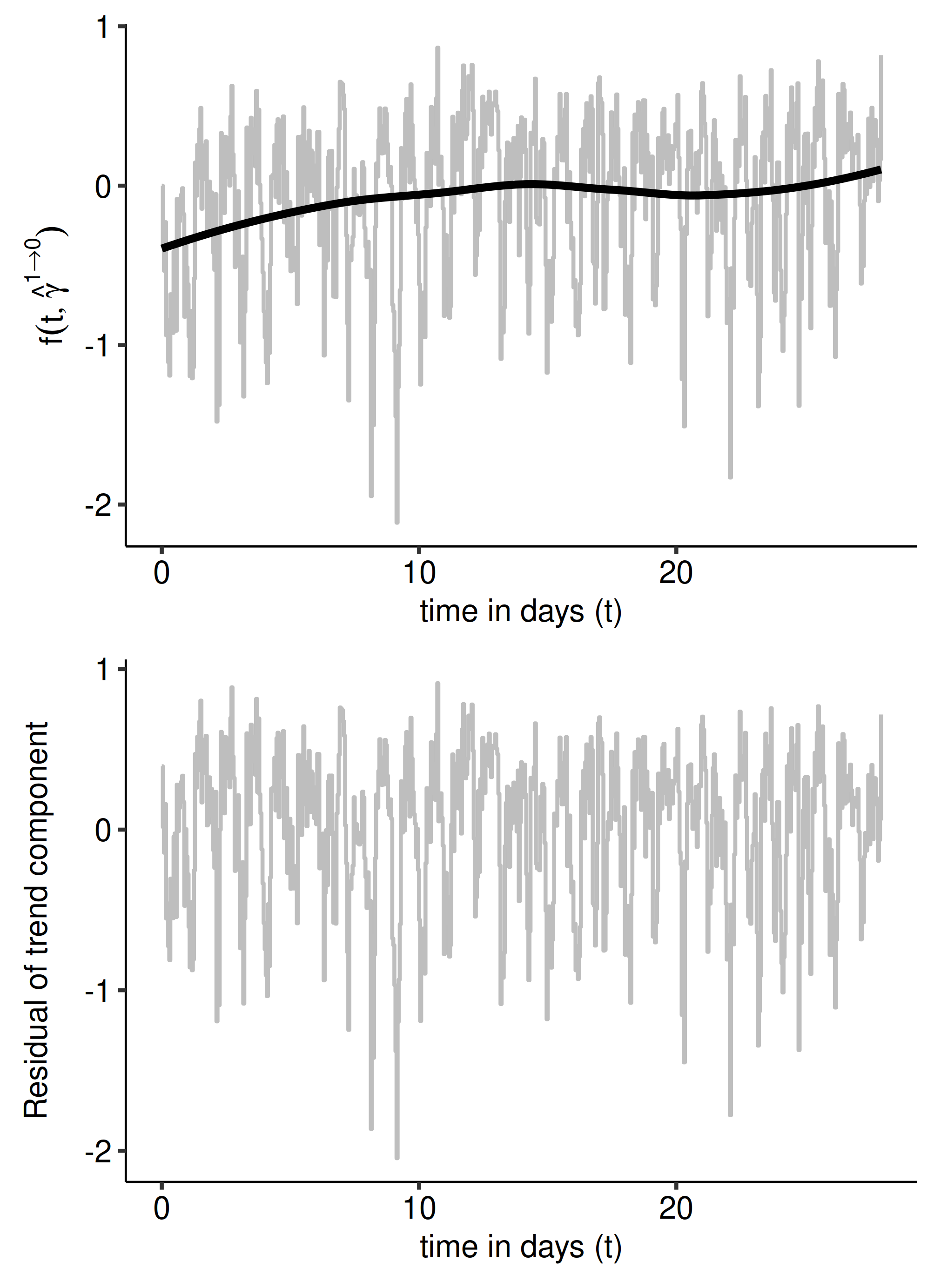}  
    \caption{\hangindent=0em Duration}
    \label{fig:subfig2}
  \end{subfigure}
    \caption{Estimates of baseline intensity of the co-location model. }
  \label{fig:est_base_proximity}
\end{figure}

\begin{figure}[t]
  \centering
  \begin{subfigure}[c]{0.48\textwidth}   
    \centering
    \includegraphics[width=\textwidth]{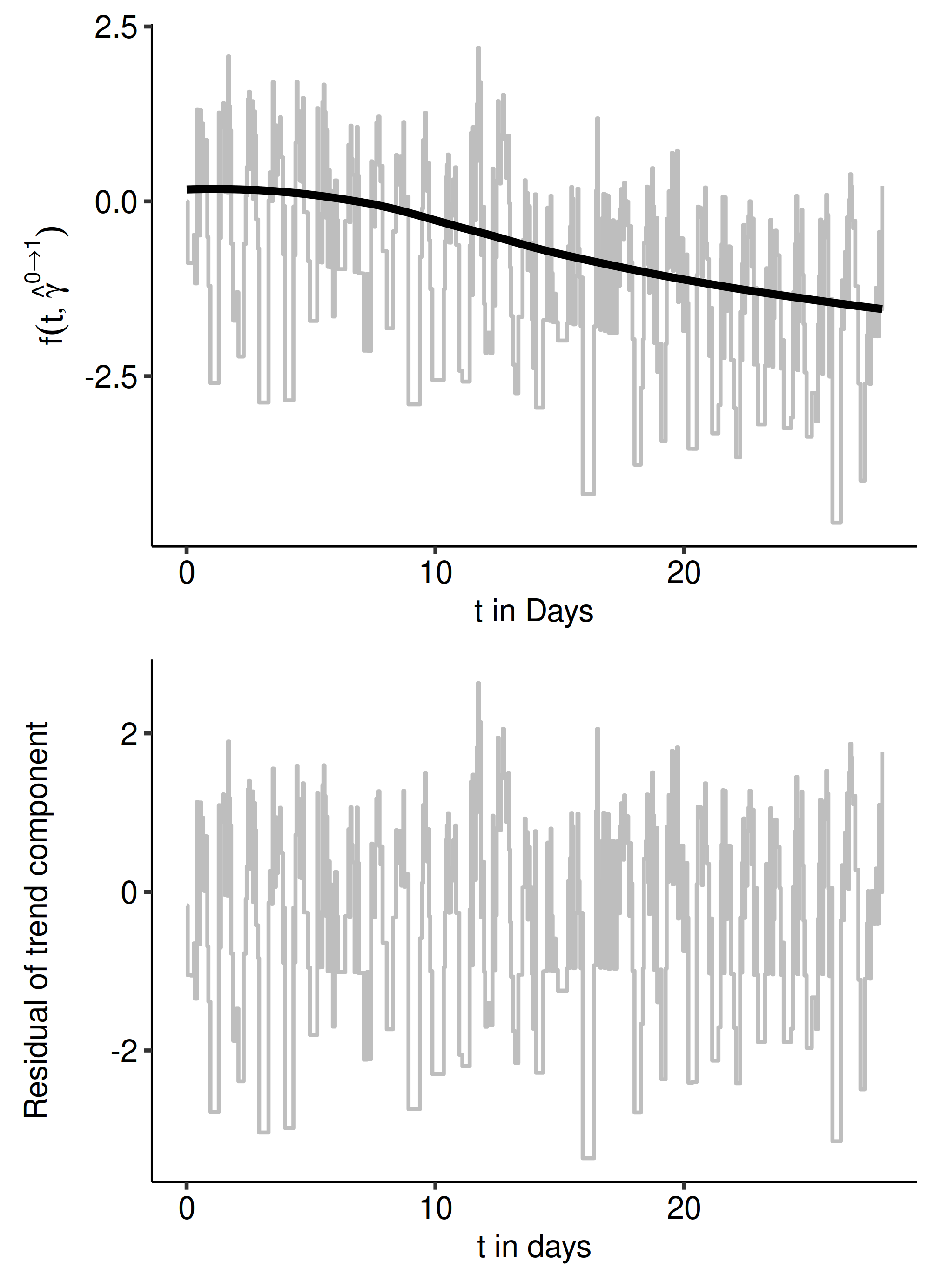} 
    \caption{\hangindent=0em Incidence}
    \label{fig:subfig1}
  \end{subfigure}
  \begin{subfigure}[c]{0.48\textwidth}  
    \centering
    \includegraphics[width=\textwidth]{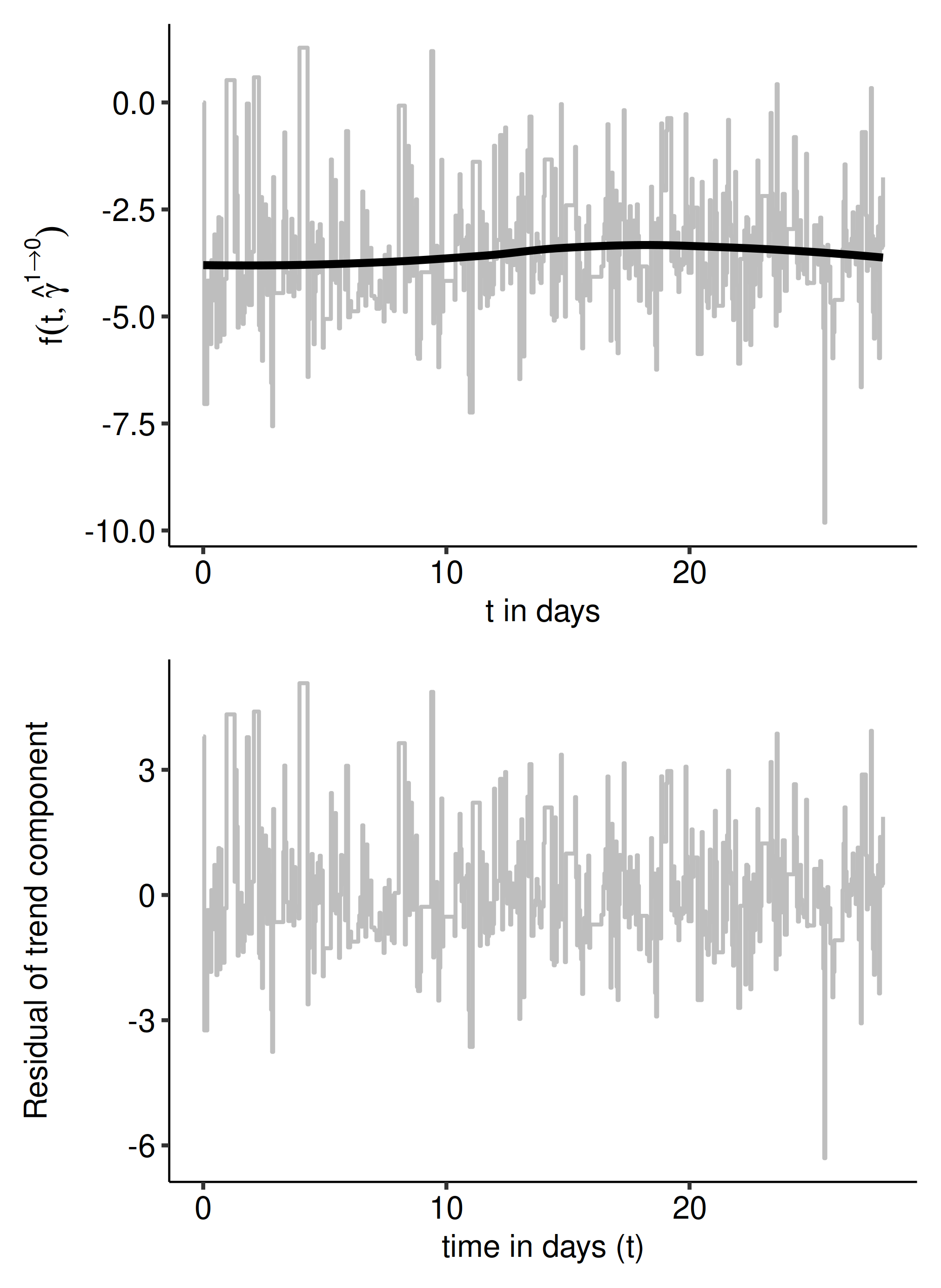}  
    \caption{\hangindent=0em Duration}
    \label{fig:subfig2}
  \end{subfigure}
    \caption{Estimates of baseline step function of the call model.}
  \label{fig:est_base_call}
\end{figure}

The estimates of the baseline step functions  $f(t,\hat {\bm  \gamma}^{\, 0\rightarrow 1})$  and  $f(t,\hat  {\bm  \gamma}^{\, 1 \rightarrow 0})$  for the co-location and call data are presented in the first rows of Figures \ref{fig:est_base_proximity} and \ref{fig:est_base_call}, respectively. 
As described in Section 5, each step in these step functions corresponds to a one-hour interval.
A clear daily cyclical pattern is evident in Figures \ref{fig:est_base_proximity} and \ref{fig:est_base_call}. To separate long-term trends from cyclical variations, we apply the local smoothing technique developed by \citetsupp{cleveland1979} to each baseline step function independently. 
The resulting trend estimates are represented by black smooth lines in all plots.

From the first row of Figures \ref{fig:est_base_proximity} and \ref{fig:est_base_call}, we observe that controlling for all other effects, the baseline step function for the initiation of a durational event shows a consistent decline over the observed period. In contrast, the baseline step function for terminating a durational event exhibits a slight increase or remains stable.
The second row of these figures displays the residuals from the local regression, which capture cyclical effects. Notably, the step function residuals for the incidence models reveal strong daily patterns, where high intensities align with daylight hours and low intensities correspond to nighttime. Additionally, in the first plot of the second row in Figure \ref{fig:est_base_proximity}, an apparent disruption in this pattern around day 15 may be associated with weekend activities.

\subsection{Model Assessment}
\label{sec:model_assess_sub}

\begin{table}[t!]
\caption{Parameter estimates $(\hat{\bm{\alpha}})$ and standard errors (SE) of the Durational Event Model applied to the co-location data.\label{tbl:results_gof_a}} 
\begin{center}
\begin{tabular}{lrrcrrcrr}
\hline
\multicolumn{1}{l}{\bfseries }&\multicolumn{2}{c}{\bfseries Full}&\multicolumn{1}{l}{\bfseries }&\multicolumn{2}{c}{\bfseries Reduced}&\multicolumn{1}{l}{\bfseries }&\multicolumn{2}{c}{\bfseries No Popularity}\tabularnewline
\cline{2-3} \cline{5-6} \cline{8-9}
\multicolumn{1}{l}{}&\multicolumn{1}{c}{$\hat{\bm{\alpha}}$}&\multicolumn{1}{c}{SE}&\multicolumn{1}{c}{}&\multicolumn{1}{c}{$\hat{\bm{\alpha}}$}&\multicolumn{1}{c}{SE}&\multicolumn{1}{c}{}&\multicolumn{1}{c}{$\hat{\bm{\alpha}}$}&\multicolumn{1}{c}{SE}\tabularnewline
\hline
{\bfseries Incidence ($\hat \alpha^{0\rightarrow 1}$)}&&&&&&&&\tabularnewline
~~Current Common Partner&2.841&.007&&&&&.786&.002\tabularnewline
~~General Common Partner&.727&.007&&&&&.041&< .001\tabularnewline
~~Number Interaction&1.139&.005&&&&&.061&< .001\tabularnewline
~~Friendship Match&.421&.011&&2.523&.011&&1.332&.011\tabularnewline
~~Gender Match&$-$.027&.011&&.220&.010&&.094&.009\tabularnewline
\hline
{\bfseries Duration ($\hat \alpha^{1\rightarrow 0}$)}&&&&&&&&\tabularnewline
~~Current Interaction&$-$.051&.001&&&&&.000&< .001\tabularnewline
~~Number Interaction&$-$.155&.005&&&&&$-$.009&< .001\tabularnewline
~~Current Common Partner&$-$.321&.007&&&&&$-$.121&.002\tabularnewline
~~General Common Partner&.098&.009&&&&&.003&< .001\tabularnewline
~~Friendship Match&$-$.574&.011&&$-$.701&.010&&$-$.539&.010\tabularnewline
~~Gender Match&$-$.019&.011&&$-$.034&.011&&$-$.028&.009\tabularnewline
\hline
~~AIC&\multicolumn{2}{c}{2,608,282}&&\multicolumn{2}{c}{2,857,341}&&\multicolumn{2}{c}{2,740,074}\tabularnewline
\hline
\end{tabular}\end{center}
\end{table}
\hide{

\alert{NEED TO UPDATE THE ESTIMATES}
\begin{table}[t!]
\caption{Parameter estimates $(\hat{\bm{\alpha}})$ and standard errors (SE) of the Durational Event Model applied to the co-location data.\label{tbl:results_gof_a}} 
\begin{center}
\begin{tabular}{lrrcrrcrr}
\hline
\multicolumn{1}{l}{\bfseries }&\multicolumn{2}{c}{\bfseries Full}&\multicolumn{1}{l}{\bfseries }&\multicolumn{2}{c}{\bfseries Reduced}&\multicolumn{1}{l}{\bfseries }&\multicolumn{2}{c}{\bfseries No Popularity}\tabularnewline
\cline{2-3} \cline{5-6} \cline{8-9}
\multicolumn{1}{l}{}&\multicolumn{1}{c}{$\hat{\bm{\alpha}}$}&\multicolumn{1}{c}{SE}&\multicolumn{1}{c}{}&\multicolumn{1}{c}{$\hat{\bm{\alpha}}$}&\multicolumn{1}{c}{SE}&\multicolumn{1}{c}{}&\multicolumn{1}{c}{$\hat{\bm{\alpha}}$}&\multicolumn{1}{c}{SE}\tabularnewline
\hline
{\bfseries Incidence ($\hat \alpha^{0\rightarrow 1}$)}&&&&&&&&\tabularnewline
~~Current Common Partner&2.730&.007&&&&&1.873&.008\tabularnewline
~~General Common Partner&.638&.008&&&&&.085&.003\tabularnewline
~~Number Interaction&1.153&.007&&&&&1.839&.005\tabularnewline
~~Friendship Match&.504&.013&&2.361&.013&&$-$5.557&.009\tabularnewline
~~Gender Match&$-$.014&.012&&.207&.012&&$-$8.692&.011\tabularnewline
\hline
{\bfseries Duration ($\hat \alpha^{1\rightarrow 0}$)}&&&&&&&&\tabularnewline
~~Current Interaction&$-$.073&.002&&&&&$-$.073&.002\tabularnewline
~~Number Interaction&$-$.236&.007&&&&&$-$.236&.007\tabularnewline
~~Current Common Partner&$-$.475&.008&&&&&$-$.475&.008\tabularnewline
~~General Common Partner&.157&.011&&&&&.158&.011\tabularnewline
~~Friendship Match&$-$.608&.013&&$-$.730&.012&&$-$.608&.013\tabularnewline
~~Gender Match&$-$.024&.013&&$-$.046&.013&&$-$.024&.013\tabularnewline
\hline
~~AIC& \multicolumn{2}{c}{$4.068 \times 10^5$}&&\multicolumn{2}{c}{$5.34 \times 10^5$} &&\multicolumn{2}{c}{$1.9 \times 10^8$}\tabularnewline
\hline
\end{tabular}\end{center}
\end{table}
}
Complementing the out-of-sample validation presented in the primary manuscript, Tables \ref{tbl:results_gof_a} and \ref{tbl:results_gof_b} provide parameter estimates derived from the subsample comprising the initial $80\%$ of observed events. 
The results of the full model specification are consistent with those of the main manuscript, providing evidence that omitting $20\%$ of the events does not affect the estimates much. 
Further, we report the AIC value for each specification in the final rows of Tables \ref{tbl:results_gof_a} and \ref{tbl:results_gof_b}. 
Since lower values indicate better model fit, the results are consistent with the out-of-sample predictions reported in the main manuscript. 
Therefore, including popularity estimates and endogenous statistics improves model fit. 
\hide{\begin{table}[t!]
\caption{Parameter estimates $(\hat{\bm{\alpha}})$ and standard errors (SE) of the Durational Event Model applied to the call data.\label{tbl:results_gof_b}} 
\begin{center}
\begin{tabular}{lrrcrrcrr}
\hline
\multicolumn{1}{l}{\bfseries }&\multicolumn{2}{c}{\bfseries Full}&\multicolumn{1}{l}{\bfseries }&\multicolumn{2}{c}{\bfseries Reduced}&\multicolumn{1}{l}{\bfseries }&\multicolumn{2}{c}{\bfseries No Popularity}\tabularnewline
\cline{2-3} \cline{5-6} \cline{8-9}
\multicolumn{1}{l}{}&\multicolumn{1}{c}{$\hat{\bm{\alpha}}$}&\multicolumn{1}{c}{SE}&\multicolumn{1}{c}{}&\multicolumn{1}{c}{$\hat{\bm{\alpha}}$}&\multicolumn{1}{c}{SE}&\multicolumn{1}{c}{}&\multicolumn{1}{c}{$\hat{\bm{\alpha}}$}&\multicolumn{1}{c}{SE}\tabularnewline
\hline
{\bfseries Incidence ($\hat \alpha^{0\rightarrow 1}$)}&&&&&&&&\tabularnewline
~~Number Interaction&1.642&.045&&&&&5.179&.026\tabularnewline
~~General Common Partner&.143&.147&&&&&2.013&.154\tabularnewline
~~Friendship Match&5.882&.134&&7.436&.144&&$-$8.195&.032\tabularnewline
~~Gender Match&.239&.091&&.103&.075&&$-$12.029&.060\tabularnewline
\hline
{\bfseries Duration ($\hat \alpha^{1\rightarrow 0}$)}&&&&&&&&\tabularnewline
~~Current Interaction&$-$.025&.028&&&&&$-$.024&.028\tabularnewline
~~Number Interaction&$-$.282&.082&&&&&$-$.303&.082\tabularnewline
~~General Common Partner&.560&.216&&&&&.546&.217\tabularnewline
~~Friendship Match&.570&.678&&.506&.671&&$-$.287&.640\tabularnewline
~~Gender Match&$-$.137&.338&&$-$.015&.349&&$-$.377&.338\tabularnewline
\hline
~~AIC& \multicolumn{2}{c}{$2.195 \times 10^4$}&&\multicolumn{2}{c}{$2.466 \times 10^4$}&&\multicolumn{2}{c}{$1.892 \times 10^8$}\tabularnewline
\hline
\end{tabular}\end{center}
\end{table}}

\begin{table}[t!]
\caption{Parameter estimates $(\hat{\bm{\alpha}})$ and standard errors (SE) of the Durational Event Model applied to the call data.\label{tbl:results_gof_b}} 
\begin{center}
\begin{tabular}{lrrcrrcrr}
\hline
\multicolumn{1}{l}{\bfseries }&\multicolumn{2}{c}{\bfseries Full}&\multicolumn{1}{l}{\bfseries }&\multicolumn{2}{c}{\bfseries Reduced}&\multicolumn{1}{l}{\bfseries }&\multicolumn{2}{c}{\bfseries No Popularity}\tabularnewline
\cline{2-3} \cline{5-6} \cline{8-9}
\multicolumn{1}{l}{}&\multicolumn{1}{c}{$\hat{\bm{\alpha}}$}&\multicolumn{1}{c}{SE}&\multicolumn{1}{c}{}&\multicolumn{1}{c}{$\hat{\bm{\alpha}}$}&\multicolumn{1}{c}{SE}&\multicolumn{1}{c}{}&\multicolumn{1}{c}{$\hat{\bm{\alpha}}$}&\multicolumn{1}{c}{SE}\tabularnewline
\hline
{\bfseries Incidence ($\hat \alpha^{0\rightarrow 1}$)}&&&&&&&&\tabularnewline
~~Number Interaction&1.640&.045&&&&&.089&.002\tabularnewline
~~General Common Partner&.144&.147&&&&&1.036&.050\tabularnewline
~~Friendship Match&5.901&.135&&7.477&.146&&5.733&.097\tabularnewline
~~Gender Match&.244&.091&&.110&.075&&.039&.055\tabularnewline
\hline
{\bfseries Duration ($\hat \alpha^{1\rightarrow 0}$)}&&&&&&&&\tabularnewline
~~Current Interaction&$-$.001&.025&&&&&$-$.002&< .001\tabularnewline
~~Number Interaction&$-$.304&.082&&&&&$-$.005&.003\tabularnewline
~~General Common Partner&.543&.217&&&&&.106&.074\tabularnewline
~~Friendship Match&.153&.658&&$-$.112&.641&&.329&.123\tabularnewline
~~Gender Match&$-$.327&.337&&$-$.260&.349&&.235&.074\tabularnewline
\hline
~~AIC&\multicolumn{2}{c}{64,343}&&\multicolumn{2}{c}{66,501}&&\multicolumn{2}{c}{69,691}\tabularnewline
\hline
\end{tabular}\end{center}
\end{table}

\subsection{Sensitivity Analysis of Change-Point Specification}
\label{sec:sensitivity}

\begin{figure}[t!]
    \centering
    \includegraphics[trim=50 0 0 0, clip, width=\textwidth]{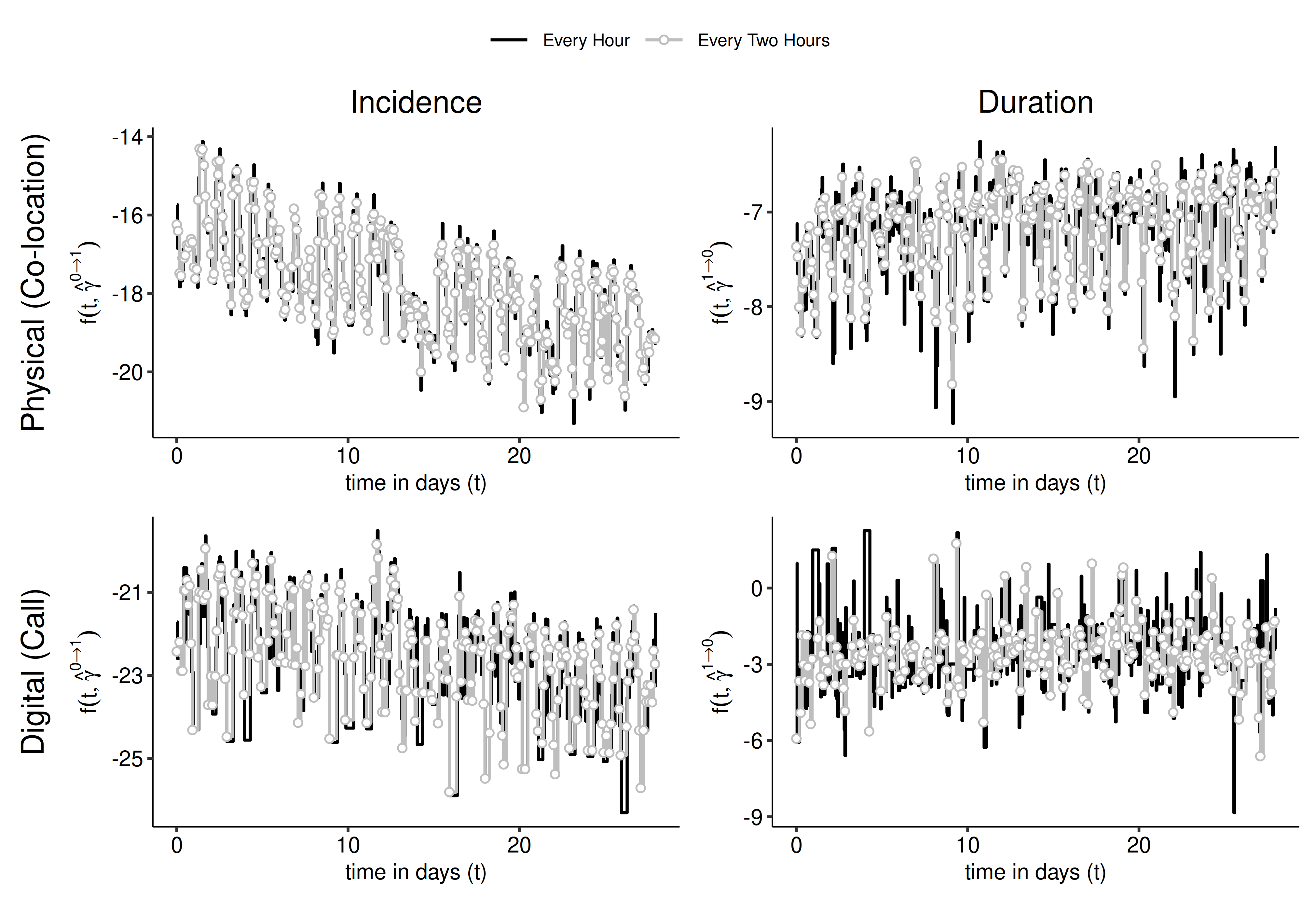}
     \caption{Comparison of the estimated baseline step function under the assumption that it can change every hour and every two hours for durational events representing co-location (first row) and call events (second row) for the incidence (first column) and duration model (second column).}
    \label{fig:baseline_sens}
\end{figure}

As detailed in the main text, the user must specify the set of change points  $0 = c_0 < c_1 < \dots < c_Q$, at which the baseline step function
$f(t, {\bm  \gamma}) = \sum_{q = 1}^{Q} \gamma_q \, \mathbb{I}(c_{q-1}\leq t < c_q)$
is allowed to change. In the application presented in Section 5, we construct an hourly grid spanning the entire 27-day observation period.
As a sensitivity check, we re-estimate the same models reported in Table 2 of the main paper, but under a coarser temporal resolution, allowing for a change point every two hours instead of one. 
The corresponding estimates of  $\hat{\bm \alpha}^{0\rightarrow 1}$  and  $\hat{\bm \alpha}^{1\rightarrow 0}$  under this specification are presented in Tables \ref{tbl:results_sens_proximity} and \ref{tbl:results_sens_call}, while the estimated baseline step function  $f(t,\hat {\bm \gamma}^{\, 0\rightarrow 1})$  and  $f(t,\hat { \bm\gamma} ^{\,1 \rightarrow 0})$  are visualized in Figure \ref{fig:baseline_sens}.
Note that the reference for the baseline step functions differs between the model with one step every one and two hours. 
To make the baseline functions comparable with one another, we add two times the average respective popularity effects to each estimate. 
All results confirm the robustness of our findings with respect to the choice of grid resolution for change points.
The estimated effects remain qualitatively the same, demonstrating that the model’s conclusions are not sensitive to the specific granularity of the baseline step function.

\begin{table}[t!]
\caption{Parameter estimates $(\hat{\bm{\alpha}})$ and standard errors (SE) of the Durational Event Model applied to the Physical (Co-location) data assuming a baseline intensity that can change every hour (left column) and every second hour (right column). \label{tbl:results_sens_proximity}} 
\begin{center}
\begin{tabular}{lrrcrr}
\hline
\multicolumn{1}{l}{\bfseries }&\multicolumn{2}{c}{\bfseries Every Hour}&\multicolumn{1}{l}{\bfseries }&\multicolumn{2}{c}{\bfseries Every two hours}\tabularnewline
\cline{2-3} \cline{5-6}
\multicolumn{1}{l}{}&\multicolumn{1}{c}{$\hat{\bm{\alpha}}$}&\multicolumn{1}{c}{SE}&\multicolumn{1}{c}{}&\multicolumn{1}{c}{$\hat{\bm{\alpha}}$}&\multicolumn{1}{c}{SE}\tabularnewline
\hline
{\bfseries Incidence ($\hat \alpha^{0\rightarrow 1}$)}&&&&&\tabularnewline
~~Current Common Partner&2.867&.006&&2.834&.006\tabularnewline
~~General Common Partner&.726&.007&&.727&.007\tabularnewline
~~Number Interaction&1.129&.005&&1.127&.005\tabularnewline
~~Friendship Match&.383&.010&&.383&.010\tabularnewline
~~Gender Match&$-$.023&.010&&$-$.021&.010\tabularnewline
\hline
{\bfseries Duration ($\hat \alpha^{1\rightarrow 0}$)}&&&&&\tabularnewline
~~Current Interaction&$-$.052&.001&&$-$.053&.001\tabularnewline
~~Number Interaction&$-$.154&.005&&$-$.160&.005\tabularnewline
~~Current Common Partner&$-$.318&.006&&$-$.328&.006\tabularnewline
~~General Common Partner&.075&.009&&.079&.009\tabularnewline
~~Friendship Match&$-$.556&.010&&$-$.554&.010\tabularnewline
~~Gender Match&$-$.017&.010&&$-$.016&.010\tabularnewline
\hline
\end{tabular}\end{center}
\end{table}

\begin{table}[t!]
\caption{Parameter estimates $(\hat{\bm{\alpha}})$ and standard errors (SE) of the Durational Event Model applied to the call data assuming a baseline intensity that can change every hour (left column) and every second hour (right column). \label{tbl:results_sens_call}} 
\begin{center}
\begin{tabular}{lrrcrr}
\hline
\multicolumn{1}{l}{\bfseries }&\multicolumn{2}{c}{\bfseries Every Hour}&\multicolumn{1}{l}{\bfseries }&\multicolumn{2}{c}{\bfseries Every two hours}\tabularnewline
\cline{2-3} \cline{5-6}
\multicolumn{1}{l}{}&\multicolumn{1}{c}{$\hat{\bm{\alpha}}$}&\multicolumn{1}{c}{SE}&\multicolumn{1}{c}{}&\multicolumn{1}{c}{$\hat{\bm{\alpha}}$}&\multicolumn{1}{c}{SE}\tabularnewline
\hline
{\bfseries Incidence ($\hat \alpha^{0\rightarrow 1}$)}&&&&&\tabularnewline
~~Number Interaction&1.630&.039&&1.630&.039\tabularnewline
~~General Common Partner&.224&.125&&.226&.121\tabularnewline
~~Friendship Match&5.701&.117&&5.706&.117\tabularnewline
~~Gender Match&.207&.085&&.209&.084\tabularnewline
\hline
{\bfseries Duration ($\hat \alpha^{1\rightarrow 0}$)}&&&&&\tabularnewline
~~Current Interaction&$-$.028&.022&&$-$.044&.020\tabularnewline
~~Number Interaction&$-$.292&.073&&$-$.164&.065\tabularnewline
~~General Common Partner&.513&.186&&.575&.170\tabularnewline
~~Friendship Match&$-$.498&.458&&$-$.660&.391\tabularnewline
~~Gender Match&$-$.324&.318&&$-$.295&.291\tabularnewline
\hline
\end{tabular}\end{center}
\end{table}

\hide{
\begin{table}[t!]
\caption{Parameter estimates $(\hat{\bm \alpha})$ and standard errors (SE) of the Durational Event Model applied to the call data assuming a baseline step function that can change every hour (left column) and every second hour (right column). \label{tbl:results_sens_call}} 
\begin{center}
\begin{tabular}{lrrcrr}
\hline
\multicolumn{1}{l}{}&\multicolumn{2}{c}{\bfseries Every Hour}&\multicolumn{1}{l}{\bfseries }&\multicolumn{2}{c}{\bfseries Every two hours}\tabularnewline
\cline{2-3} \cline{5-6}
\multicolumn{1}{l}{\textbf{Summary Statistic}}&\multicolumn{1}{c}{$\hat{\bm \alpha}$}&\multicolumn{1}{c}{SE}&\multicolumn{1}{c}{}&\multicolumn{1}{c}{$\hat{\bm \alpha}$}&\multicolumn{1}{c}{SE}\tabularnewline
\hline
{ Incidence ($\hat {\bm \alpha}^{0\rightarrow 1}$)}&&&&&\tabularnewline
~~General Common Partner&1.631&.039&&1.631&.039\tabularnewline
~~Number Interaction&.224&.125&&.227&.121\tabularnewline
~~Friendship Match&5.687&.116&&5.681&.116\tabularnewline
~~Gender Match&.203&.085&&.201&.084\tabularnewline
\hline
{ Duration ($\hat {\bm \alpha}^{1\rightarrow 0}$)}&&&&&\tabularnewline
~~Current Interaction&$-$.053&.024&&$-$.060&.023\tabularnewline
~~Number Interaction&$-$.274&.073&&$-$.162&.065\tabularnewline
~~General Common Partner&.518&.186&&.579&.170\tabularnewline
~~Friendship Match&$-$.337&.464&&$-$.670&.390\tabularnewline
~~Gender Match&$-$.227&.320&&$-$.211&.287\tabularnewline
\hline
\end{tabular}\end{center}
\end{table}
}

\subsection{REM Application to Texting}
\label{sec:rem}
For the application to the texting data between the students, we apply similar inclusion criteria as in the applications shown in the main analysis. 
However, we apply the condition that all included actors need to have at least one relational event, i.e., texting event. 
Thereby, we obtain $\mbox{38,286}$ relational events between $N = 426$ students. 
Each relational event $d = (i,j,t)$ is a tuple representing a text message between students $i$ and $j$ at time $t$.

In accordance with (2), the intensity of an event at time $t \in \mathscr{T}$ for pair $(i,j)\in \dyadset$ is:
\be
\label{eq:intensity_rem_conditional}
      \lambda_{i,j}\left(t\mid \mathscr{H}_t, \bm{\theta}\right) &=& \exp\left(\bm \alpha^\top\bm s_{i,j}(\mathscr{H}_t) + \beta_i +\beta_j + f(t,\bm{\gamma})\right).
\ee
The summary statistics are 
\beno 
\bm s_{i,j}(\mathscr{H}_t)
&=&
\left(
\begin{array}{c}
 \log\left(\sum_{h \notin \{i,j\}} v_{i,h}(t^-)\, v_{h,j}(t^-)+1\right) \\
 \log(N_{i,j}(t^-)+1) \\
  \mathbb{I}(x_{i,j,1} = 1)  \\
  \mathbb{I}(x_{i,2} = x_{j,2})
\end{array}
\right),
\ee
 where $v_{i,j}(t^-)$ indicates whether actors $i$ and $j$ have ever interacted until time $t^-$, which is the time right before $t$.
The covariates are defined in Section 5.   


 \begin{table}[t!]
\caption{Parameter estimates $(\hat{\bm{\alpha}})$ and standard errors (SE) of the Relational Event Model applied to the Texting data. \label{tbl:results_sms}} 
\begin{center}
\begin{tabular}{lrrr}
\hline
\multicolumn{1}{l}{}&\multicolumn{1}{c}{$\hat{\bm{\alpha}}$}&\multicolumn{1}{c}{SE}&\multicolumn{1}{c}{$2^{\hat{\bm{\alpha}}}$}\tabularnewline
\hline
Common Partner&.304&.052&1.235\tabularnewline
Number Interaction&2.230&.011&4.691\tabularnewline
Friendship Match&$-$.412&.134&\tabularnewline
Gender Match&.228&.039&\tabularnewline
\hline
\end{tabular}\end{center}
\end{table}

 \paragraph*{Summary Statistics.}
 The estimates of $\bm{\alpha}$ are given in Table \ref{tbl:results_sms} and can be interpreted as detailed in Section 2.2. 
 The first common partner and joint interaction have a multiplicative effect of $1.235$ and $4.691$ on the intensity between students $i$ and $j$. 
 These two coefficients demonstrate that sharing a common text partner incentivizes triadic closure and that students repeatedly exchange texts between them. 
 Contrary to these findings, being friends on Facebook has a small but significant negative effect on the intensity. 
 Finally, we observe a positive gender-specific homophily effect, whereby two students of the same gender are more likely to exchange text messages than mixed gender pairs.

\begin{figure}[t!]
    \centering
    \includegraphics[trim=300 300 250 250, clip,width=0.7\textwidth]{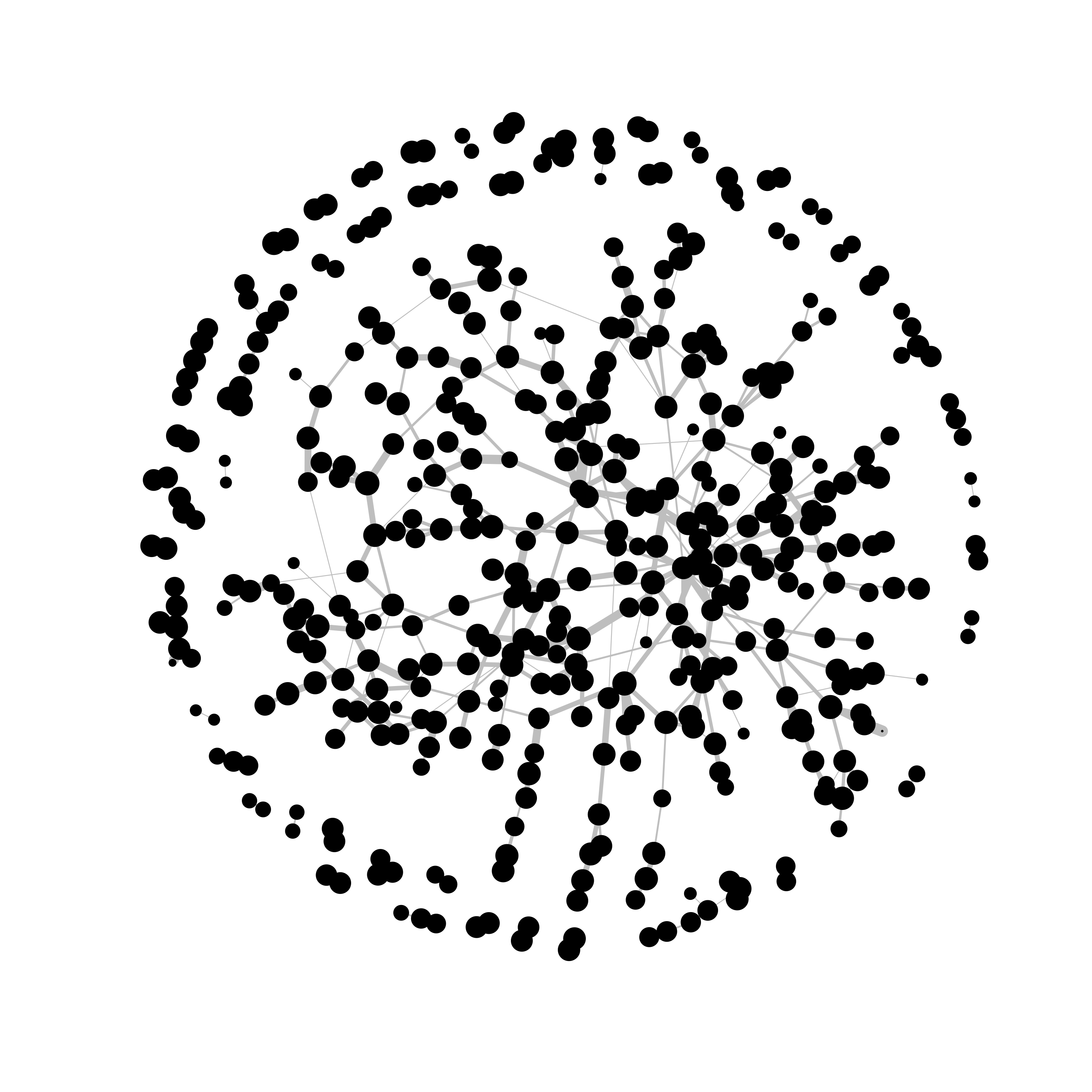}
     \caption{Observed texting events over the studied 28 days. The number of text messages between particular students is represented by the thickness of the edge and the size of the nodes represents the relative popularity estimate.}
    \label{fig:network}
\end{figure}

\paragraph*{Popularity Estimates.}
We visualize the observed dyadic text message exchanges in Figure \ref{fig:network} as a weighted network, where the weight of each edge represents the logarithm of the number of SMS events between the corresponding pair of students. 
We used the Fruchterman-Reingold algorithm \citepsupp{fruchterman1991} for this visualization. 
The size of each node is scaled relative to the corresponding popularity estimate  $\hat{\beta}_i$.
Figure \ref{fig:network} reveals that students with higher popularity estimates do not necessarily occupy central positions in the network. 
One possible explanation is that the interaction patterns between actors in central regions of the network are well captured by the summary statistics without the need for additional popularity effects.


\begin{figure}[t!]
    \centering
    \includegraphics[width=0.8\textwidth]{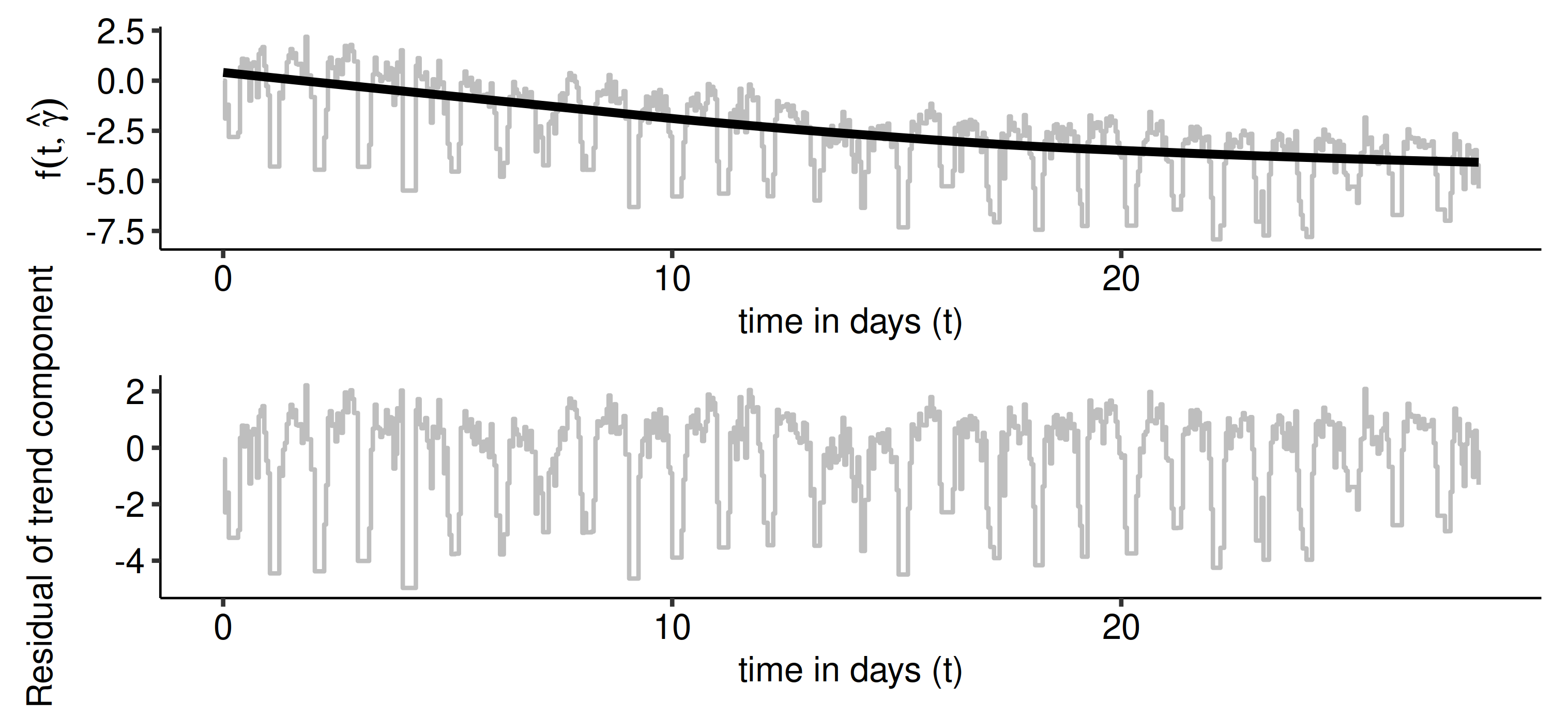}
     \caption{Estimated baseline step function for the REM applied to the Texting Data.}
    \label{fig:baseline_sms}
\end{figure}

\paragraph*{Baseline Step Function.}
Finally, we visualize the estimated baseline step function  $f(t,\hat{\bm{\gamma}})$  in Figure \ref{fig:baseline_sms}. 
Consistent with the patterns observed in Figures \ref{fig:est_base_proximity} and \ref{fig:est_base_call}, the function exhibits distinct daily cycles alongside a gradual downward trend over time. 

\newpage 
\bibliographystylesupp{chicago}
\bibliographysupp{references}

\end{document}